\documentclass[11pt, usenames,dvipsnames]{article}
\pdfoutput=1

\usepackage{bm,wrapfig,float,array,mathtools}
\usepackage{amsfonts}
\usepackage{amssymb}
\usepackage{cite}
\usepackage{amsmath}
\usepackage{color}
\usepackage{graphicx}
\usepackage{hyperref}
\usepackage{float}
\usepackage[T1]{fontenc}
\usepackage[utf8]{inputenc}
\usepackage{alphabeta}
\usepackage{fancyvrb}
\usepackage[dvipsnames]{xcolor}
\usepackage{bold-extra}
\usepackage{lmodern}
\usepackage{cleveref}
\usepackage[normalem]{ulem}
\usepackage{tikz-cd}
\newcommand{\RNum}[1]{\uppercase\expandafter{\romannumeral #1\relax}}
\usepackage[dvipsnames]{xcolor}
\usepackage[framemethod=TikZ]{mdframed}
\usepackage{amsthm}
\usepackage{bbold}

\usepackage{multirow}

\usepackage{subfigure}
\usepackage{paralist}
\usepackage{tikz}
\usepackage{diagbox}
\usetikzlibrary{positioning}
\usetikzlibrary{calc}
\usetikzlibrary{decorations.pathreplacing,calligraphy}

\usepackage{xstring}
\usetikzlibrary{decorations.pathmorphing} 
\usetikzlibrary{decorations.markings} 
\usetikzlibrary{arrows} 
\usetikzlibrary{shapes} 
\usetikzlibrary{matrix} 
\usetikzlibrary{positioning} 
\usepackage[english]{babel} 
\usepackage[autostyle]{csquotes}
\usetikzlibrary{shapes.multipart}
\usetikzlibrary{patterns}

\newcommand{\cL}{\mathcal{L}}

\tikzstyle{brane}=[draw]
\tikzset{D7/.style={circle, draw=black, inner sep=0pt, fill=white, minimum size=3mm}}
\tikzset{hasse/.style={circle, fill,inner sep=2pt}}
\tikzset{flavor/.style={regular polygon,fill=white,regular polygon sides=4,inner sep=2.5pt, draw}}
\tikzset{gauge/.style={circle, draw,inner sep=2.5pt}}
\tikzset{gaugeb/.style={circle, draw,fill=black,inner sep=2.5pt}}
\tikzset{gauger/.style={circle, draw,fill=cyan,inner sep=2.5pt}}
\tikzset{gaugeg/.style={circle, draw,fill=red,inner sep=2.5pt}}
\tikzset{SUd/.style={circle, draw=black, inner sep=0pt, fill=yellow, minimum size=2mm}}
\tikzset{bd/.style={circle, draw=black, inner sep=0pt, fill=black, minimum size=2mm}}
\tikzset{wd/.style={circle, draw=black, inner sep=0pt, fill=white, minimum size=2mm}}
\tikzset{Dynkin/.style={circle, draw=black, inner sep=0pt, fill=white, minimum size=2mm}}
\tikzstyle{ligne}=[draw, thick] 
\tikzset{doublearrow/.style={ draw=black!75, color=black!75, thick, double distance=3pt, }}

\numberwithin{equation}{section}  

\graphicspath{ {figs/} }

\newcommand{\be}{\begin{equation}}
\newcommand{\ee}{\end{equation}}
\newcommand{\ba}{\begin{aligned}}
\newcommand{\ea}{\end{aligned}}

\newcommand{\R}{\mathbb{R}}

\begin{document}

\baselineskip=18pt  
\numberwithin{equation}{section}  
\allowdisplaybreaks  


%
%


\thispagestyle{empty}

\vspace*{0.8cm} 
\begin{center}
{\huge The Line, the Strip and the Duality Defect}\\

 \vspace*{1.5cm}
{\large Francesco Bedogna$^{1}$, Salvo Mancani$^{1}$} \
 \vspace*{.2cm}
 \smallskip

{\it $^1$ Dipartimento di Fisica e Astronomia “Galileo Galilei”, Università di Padova,\\ Via Marzolo 8, 35131 Padova, Italy  }\\

\smallskip

{\it $^1$ INFN, Sezione di Padova Via Marzolo 8, 35131 Padova, Italy   }\\

\vspace*{2cm}
\end{center}

\noindent

In the Symmetry Topological Field Theories (SymTFT) that describes the exotic models XY-plaquette and XYZ-cube, we construct codim-1 condensation defects by higher gauging with discrete torsion the non-compact symmetry of the bulk. In the framework of SymTFT \textit{Mille-feuille}, which captures the Lorentz-invariance breaking subsystem symmetries, these models are dual to foliated versions of Maxwell theory. We show first that the XY-plaquette model admits a $\theta$-term. Then, we show these condensation defects realize non-invertible self-duality symmetries at any value of the coupling. In the XYZ-cube model such symmetry is discrete. On the other hand, we find that the XY-plaquette has a non-invertible continuous $SO(2)$ symmetry, thus extending the results in the current literature.

 \newpage

\tableofcontents


\newpage

\section{Introduction}

Since the seminal work of \cite{Gaiotto:2014kfa}, the notion of global symmetries has been generalized to a perspective that associate them to topological operators. This broader picture includes extended charged operators and non-invertible symmetries, see \cite{McGreevy:2022oyu, Schafer-Nameki:2023jdn, Shao:2023gho, Bhardwaj:2023gsy, Luo:2023lsa, Brennan:2023ggs, Costa:2024scs} for instances and review, and subsystem symmetries, which act only on submanifold of the spacetime. 

The symmetry structure of a quantum field theory (QFT) can be encoded into topological information, in the framework of the Symmetry Topological Field Theory (SymTFT). In this construction, a $d$-dimensional QFT with a certain symmetry $G$ is described using a $d+1$-dimensional topological field theory with gauge group $G$, coupled to the physical theory living on the boundary $\mathcal{B}^{\mathrm{phys}}$ of the bulk. The symmetry operators and the charged operators of the field theory are realized as the topological operators of the bulk theory. A set of topological boundary conditions, encoded on a second boundary $\mathcal{B}^{\mathrm{top}}$, determines which bulk operators can be projected or can terminate on the boundary. Therefore, in the SymTFT the form of the spacetime is $M_{d+1} = M_d \times I$, where $I$ is a finite interval. By compactifying the interval, the physical theory is recovered together with its symmetry structured. Different choices of boundary conditions correspond to different global variants of the same QFT. 

SymTFTs have proven to be a powerful tool to describe both finite \cite{Gaiotto:2014kfa, Ji:2019jhk, Ji:2019eqo, Kong:2015flk, Kong:2020cie, Gaiotto:2020iye, Apruzzi:2021nmk, Freed:2022qnc, Apruzzi:2022rei, Kaidi:2022cpf, Antinucci:2022vyk, Bhardwaj:2023ayw, Kaidi:2023maf, Zhang:2023wlu, Bhardwaj:2023fca, Bartsch:2023wvv, Apruzzi:2022dlm, Schafer-Nameki:2023jdn, Cvetic:2024dzu, Baume:2023kkf} and continuous symmetries \cite{Brennan:2024fgj, Antinucci:2024zjp, Bonetti:2024cjk, Apruzzi:2024htg, Arbalestrier:2024oqg}, depending on the choice of compact or non-compact gauge fields in the bulk TFT. Within the framework of non-compact SymTFTs, the gauging of a non-compact continuous symmetry with flat connections has been used to argue for a non-invertible self-duality symmetry at any value of the coupling, not only rational. In particular, \cite{Argurio:2024ewp} constructs a self-duality symmetry for the compact boson in 2d at any value of the compactification radius $R$, and \cite{Paznokas:2025auw} constructs an $SO(2)$ duality symmetry in 4d Maxwell. In both cases, they define condensation defects \cite{Gaiotto:2019xmp, Roumpedakis:2022aik} supported on a codimension-one surface in the bulk, in order to gauge a subgroup $\mathbb{R}$ of the bulk global symmetry. While in general such condensation defects are invertible, they give rise to non-invertible fusion rules once open and thus have boundaries. Indeed, that is the case for the 2d compact boson, where T-duality defects are generated by the boundaries of these open condensation defects, at any point of the conformal manifold. Similarly, for Maxwell theory, by gauging the continuous symmetry in the bulk with discrete torsion, the boundaries of the condensation defects realize an $SO(2)$ duality symmetry at any value of the complex coupling $\tau$. These results generalize the conclusions of \cite{Niro:2022ctq}, limited at rational couplings, and, from a different perspective, they are consistent with \cite{Hasan:2024aow}. 

The SymTFT also accommodates modulated and subsystem symmetries \cite{Cao:2023rrb, Pace:2024tgk, Apruzzi:2025mdl}. Subsystem symmetries arise in fracton phases of matter with constrained mobility along sub-manifolds such as lines or planes. In such models, Lorentz-invariance is partially broken. Fracton systems have attracted significant attention due to their exotic properties, including infinite ground state degeneracy and UV/IR mixing. Encoding their symmetry structure into the SymTFT construction further broaden its validity and the perspective of generalized symmetries within a unique framework. In \cite{Cao:2023rrb}, the bulk theory consists of gauge fields living on leaves of the spacetime, encoding the subsystem transformations in a foliated nature of the field theory. Notice that the bulk theory is only partially topological on the directions orthogonal to the foliation. The works in \cite{Apruzzi:2025mdl, Ohmori:2025fuy} use such foliated bulk TFT and the known duality with exotic theories \cite{Spieler:2023wkz} to construct gapless exotic/foliated duality for various models: the XY-plaquette, XYZ-cube, $\phi$ and $\hat{\phi}$ theories \cite{Seiberg:2020bhn, Seiberg:2020cxy, Seiberg:2020wsg}. Interestingly, upon slab compactification these are dual to a foliated version of a 3d Maxwell theory, coupled with foliated fields via Chern-Simons terms. 

In this work, we construct the condensation defects that realize the 0-form symmetries of the bulk exotic and foliated theories discussed in \cite{Apruzzi:2025mdl}, in particular the models that yield, after compactification of the interval, the XY-plaquette and the XYZ-cube of \cite{Seiberg:2020bhn, Seiberg:2020cxy, Seiberg:2020wsg}. By implementing higher gauging of certain combinations of $\mathbb{R}$ symmetries with discrete torsion, we generate the $SL(2,\mathbb{R})$ of the 3+1 bulk and the discrete symmetries of the 4+1 bulk, respectively the SymTFT of the XY and XYZ models. A subset of these generators leaves the physical boundary conditions invariant. We find that the XY-plaquette has an $SO(2)$ non-invertible symmetry at arbitrary value of the couplings, in analogy with Maxwell theory. This result extends the non-invertible symmetry described in \cite{Spieler:2024fby}. We further observe that an exotic $\theta$-term can be added to the topological boundary, thereby enriching the space of admissible boundary conditions for bulk operators; this corresponds to the inclusion of an exotic theta term in the XY-plaquette theory. On the other hand, the XYZ-cube admits only a discrete non-invertible symmetry, due the absence of continuous bulk symmetry, in agreement with the result of \cite{Spieler:2024fby}. Let us stress that the duality symmetry defects of the physical theory are realized from the boundaries of the condensation defects in the bulk, after the bulk 0-form symmetry is gauged. In doing so, the condensation defects become transparent, while their boundaries remain as genuine operators of the physical theory that realize the duality symmetry \cite{Burbano:2021loy, Kaidi:2022cpf, Antinucci:2022vyk}. We focus in constructing the condensation defects and deriving the fusion properties of their boundaries.

In constructing the condensation defects of both models, we rely on the exotic description of the bulk, since the corresponding foliated expressions has proven to be technically complicated. This is due to the fact that the bulk operators, lines and strips, are partially topological and it is not clear how to define the homology group where they take value. Moreover, from a path integral perspective in the exotic model, one would expect a symmetry group of the bulk that contains $SL(2,\mathbb{R})$. However, this kind of rotations seems to break the foliation of the theory. In any case, because of the exotic/foliated duality we expect such condensation defects to exist in the foliated theory. We leave the problem of describing explicitly the foliated condensation defects and foliated higher gauging to future works. 

This paper is organized as follows. In sec. \ref{sec::symtft}, we review the XY-plaquette and XYZ-cube models and the gapless duality discussed in \cite{Apruzzi:2025mdl}. In sec. \ref{sec:GenDuality} we show how a different topological boundary conditions yields an exotic $\theta$-term when compactifying the XY-plaquette, and that a self-duality holds in presence of this term. In sec. \ref{sec:CondDefects}, we construct the condensation defects that implements the 0-form symmetries of the bulk of the 3+1 and 4+1 dimensional SymTFTs, and determine which of their subset can end on the physical boundary. We show that the open defects have non-invertible fusion rules, and hence they generate non-invertible $SO(2)$ for the XY-plaquette, at any value of the coupling, and a discrete non-invertible for the XYZ-cube. Finally, in \ref{sec:conclusion} we draw conclusions and highlight some interesting questions.

\section{Mille-feuille SymTFT and gapless exotic/foliated duality \label{sec::symtft}}

In this section we review the gapless exotic/foliated duality of \cite{Apruzzi:2025mdl}, and highlight the features that will be crucial in the next section. We focus on two models, XY-plaquette and XYZ-cube introduced in \cite{Seiberg:2020bhn, Seiberg:2020cxy, Seiberg:2020wsg}. Throughout the all manuscript we work in Euclidean signature.

\begin{figure}[h!]
    \centering
    \includegraphics[width=0.4\linewidth]{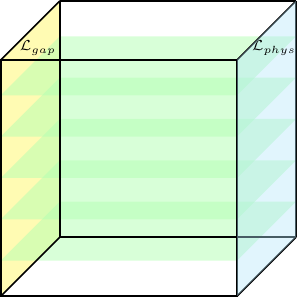}
    \caption{The Mille-feuille. The vertical direction is the foliated one. Some defects of the theory will be topological only on the green layer plane.}
\label{fig:millefoglie}
\end{figure}

\subsection{XY-plaquette model}\label{sec:XYplaquette}

The XY-plaquette model defined on 3d compact spacetime $M$ has action \cite{Seiberg:2020bhn}
\begin{equation}\label{eq::xy-Lagrangian}
    S_{xy}=\frac{1}{2\pi}\int_Md^3x \, \frac{\mu_0}{2}\left(\partial_t\phi \right)^2 + \frac{1}{2\mu}\left(\partial_x\partial_y\phi \right)^2 \; ,
\end{equation}
where $\phi$ is a compact scalar field with radius $R$ and subject to the identification $\phi\sim\phi+2\pi (n_x(x)+n_y(y))$. The couplings $(\mu_0,\mu)$ encodes information about the radius $R$ and a rescale parameter $g$. The peculiar feature of this model is that Lorentz-invariance is broken, and space rotations are broken down to a discrete $\mathbb{Z}_4$ symmetry. The scalar field transform as a spin-0 representation of $\mathbb{Z}_4$. This theory also has $U(1)$ momentum and $U(1)$ winding dipole symmetry with a mixed anomaly. 

The model has a dual description in terms of a spin-2 field $\phi^{xy}$ with action
\begin{align}\label{eq:dualXY}
    S_{xy}=\frac{1}{2\pi} \int_M d^3x \, \frac{\tilde{\mu}_0}{2}\left(\partial_t\phi^{xy} \right)^2 + \frac{1}{2\tilde{\mu}}\left(\partial_x\partial_y\phi^{xy} \right)^2 \; ,
\end{align}
with the identification
\begin{align}\label{eq:XYduality}
    \tilde{\mu}_0 &= \frac{\mu}{4 \pi^2} \; , \nonumber \\[4pt]
    \tilde{\mu} &= 4 \pi^2 \mu_0 \; ,
\end{align}
provided that the spatial rotation symmetry is composed with charge conjugation. This duality exchanges momentum and winding modes, similarly as $T$-duality for a 1+1d compact boson.

The theory and its symmetries can be described using the SymTFT with exotic bulk \cite{Apruzzi:2025mdl}
\begin{equation}\label{eq:ExoticSymTFT}
    S_{exo} = \frac{i}{2\pi}  \int_{M\times I} d^4x \left[ A_t \left( \partial_r \Tilde{A}_{xy}-\partial_x\partial_y\Tilde{A}_r \right) + A_r \left( \partial_x\partial_y\Tilde{A}_t-\partial_t\Tilde{A}_{xy} \right) + A_{xy} \left( \partial_r\Tilde{A}_t-\partial_t\Tilde{A}_r \right) \right] ,
\end{equation}
where the interval $I=[ 0, L] $ is parametrized by the radial coordinate $r$, and the gauge fields $A$ and $\tilde{A}$ take value in $\mathbb{R}$, with gauge transformations
\begin{align}\label{eq:XYexoticgauge}
        \delta A_t &=\partial_t\lambda \; , \quad \delta \Tilde{A}_t=\partial_t\Tilde{\lambda} \nonumber \\[4pt]
        \delta A_r &=\partial_r\lambda \; , \quad \delta \Tilde{A}_r=\partial_r\Tilde{\lambda} \nonumber \\
        \delta A_{xy} &=\partial_x\partial_y \lambda \; , \quad \delta \Tilde{A}_{xy}=\partial_x\partial_y\Tilde{\lambda} \; .
\end{align}

The topological boundary is placed at $r=0$ and it contains a compact scalar edge mode $\phi$ with radius $R$ and action
\begin{align}\label{eq:XYTopologicalBoundary}
    S_{0} = \frac{iR}{2\pi} \int_{M_0} d^3 x\,\phi \left( \partial_t \tilde{A}_{xy}-\partial_x\partial_y \tilde{A}_t \right)   \; ,
\end{align}
and $\delta\phi = - \lambda/R$ for gauge-invariance. The topological b.c. read
\begin{align}
    A_t &= - R \, \partial_t \phi \; , \nonumber \\[4pt]
    A_{xy} &= - R \, \partial_x \partial_y \phi \; ,
\end{align}
wheres the physical boundary lives at $r=L$ with action
\begin{align}
    S_L = \frac{1}{4\pi}\int_{M_L} d^3x\, \left[ g\left( \Tilde{A}_t \right)^2 + \left( \Tilde{A}_{xy} \right)^2 \right]\; ,
\end{align}
with b.c. 
\begin{align}\label{eq:XYconformalBoundary}
    \tilde{A}_t &= \frac{i}{g} A_{xy} \; , \nonumber \\[4pt]
    \tilde{A}_{xy} &= i A_t \; .
\end{align}
The boundary conditions set the representation under spatial $\mathbb{Z}_4$ of the bulk gauge fields. In facts, $A_t$ transforms as a spin-0 field and $A_{xy}$ as a spin-2 field, whereas the representations are exchanged for the tilde fields.

Upon interval compactification and imposing the above set of b.c., we recover the XY-plaquette theory eq. \eqref{eq::xy-Lagrangian}, identifying $R^2 = \mu_0$ and $R^2/g = 1/\mu$. The bulk of the SymTFT is dual to a foliated description whose action reads \cite{Cao:2023rrb, Apruzzi:2025mdl} 
\begin{align}\label{eq:XYFoliated}
    S_{fol} =\frac{i}{2\pi} \int_{M \times I} C^{x} \wedge dB^{x} \wedge dx + C^{y} \wedge dB^{y} \wedge dy + C^{x} \wedge b \wedge dx + C^{y} \wedge b \wedge dy + c \wedge db \; ,
\end{align}
where $b$ is a 2-form, $c$ is a 1-form and $B^{i}$, $C^{i}$ are foliated 1-forms, i.e. they live on leaves orthogonal to directions $i=x,y$. All fields are real-valued, and for gauge-invariance they transform as
\begin{equation}\label{eq:XYGaugeTransformBulk}
    \begin{split}
        &\delta b=d\chi_1 \; ,\\
        &\delta c=d\lambda_0-\sum_i\lambda^{i} dx^{i} \; ,\\
        &\delta B^{i}=d\chi_0^{i}-\chi_1 \; ,\\
        &\delta C^{i}=d\lambda^{i} \; .
    \end{split}
\end{equation}

By integrating out $C_r^{i}$, $b_{rt}$, $b_{rx}$, $b_{ry}$ from the action, it yields eq. \eqref{eq:ExoticSymTFT}. The boundary modes read
\begin{align}
    S_0 = \frac{iR}{2\pi}\int_{M_0} \Phi^{x} \wedge  (dB^{x} + b) \wedge dx + \Phi^{y} \wedge ( dB^{y} + b) \wedge dy + \phi \wedge db \; ,
\end{align}
at the topological boundary, where $\Phi^{i}$ and $\phi$ are compact scalars with gauge transformations $\delta\phi =-\lambda_0/R $ and $\delta\Phi^{i} = -\lambda^{i} / R$. On the other boundary, the action reads
\begin{align}
    S_L =& \; \frac{1}{4\pi}\int_{M_L} \left( b + dB^{x} + dB^{y} \right) \wedge*_2 \left( b + dB^{x} + dB^{y} \right) \wedge dt \nonumber \\[4pt]
    & + g \left( B^{x} - B^{y} \right)\wedge*_1 \left(B^{x} - B^{y} \right) \wedge dx \wedge dy \; ,
\end{align}
where $*_2$ and $*1$ are Hodge operators on space and time respectively. The b.c. at the gapped boundary reads
\begin{align}\label{eq:XYFoliatedbcTop}
    & c = R \left( d \phi - \sum_i \Phi^{i} dx^{i} \right) \; , \nonumber \\[5pt]
    & C^{i} \wedge dx^{i} = - R \, d\Phi^{i} \wedge dx^{i} \; ,
\end{align}
while the physical boundary enforces the b.c.
\begin{align}\label{eq:XYFoliatedbcPhys}
    & c = i *_2 \left( b + \sum_i dB^{i} \right) \wedge dt \; , \nonumber \\[5pt]
    & C^{x} \wedge dx = + i g  \left( B^{x}_t - B^{y}_t \right) dx \wedge dy \wedge dt +  i d *_2 \left( b + dB^{x} + dB^{y} \right) \wedge dt \; , \nonumber \\[5pt]
    & C^{y} \wedge dy = - i g \left( B^{x}_t - B^{y}_t \right) dx \wedge dy \wedge dt + i d *_2 \left( b + dB^{x} + dB^{y} \right) \wedge dt \; .
\end{align}
By compactifying and integrating out the scalars $\Phi^x$ and $\Phi^y$, we obtain the XY-plaquette. On the other hand, by integrating out the scalar $\phi$, we get the foliated dual theory to the XY-plaquette
\begin{align}\label{eq:FoliatedMaxwell}
    S_{\mathrm{gapless}} =& \frac{1}{2\pi}\int_{M_0} iR \left[ \Phi^{x}\wedge d( B^{x} + a)\wedge dx+\Phi^{y}\wedge d( B^{y} + a)\wedge dy \right] \nonumber \\[5pt]
    & + \frac{1}{2} d\left( a + B^{x} + B^{y} \right) \wedge*_2 d\left( a + B^{x} + B^{y} \right) \wedge dt \nonumber \\[4pt]
    & + \frac{1}{2} g \left( B^{x}-B^{y} \right)\wedge*_1 \left(B^{x}-B^{y}\right)\wedge dx\wedge dy \; ,
\end{align}
where $b = da$. This theory is a foliated version of a 3d Maxwell with gauge field $\left( a + B^{x} + B^{y} \right)$, coupled to the scalars $\Phi^x$ and $\Phi^y$.

\subsubsection*{Partially topological operators}
The set of (partially) topological operators in the bulk consists of line and strip operators, in both exotic and foliated frameworks. The lines are supported on a closed curve $\gamma$ inside the submanifold at fixed $(x,y)$, where they can be topologically deformed. In foliated fields these are
\begin{align}\label{eq:XYFoliatedLines}
    & V_{\alpha} (x,y) [\gamma] = \exp \left( i \alpha \oint_\gamma c \right) \; , \nonumber \\
    & U_{\beta} (x,y) [\gamma] = \exp \left( i \beta \oint_\gamma B^x-B^y \right) \; ,
\end{align}
while in exotic fields
\begin{align}\label{eq:XYExoticLines}
    V_{\alpha}(x,y)[\gamma] & = \exp \left( i \alpha \oint_\gamma A_t dt + A_r dr \right) \; , \nonumber \\[4pt]
    U_{\beta}(x,y)[\gamma] & = \exp \left( i \beta \oint_\gamma \Tilde{A}_t dt + \Tilde{A}_r dr \right) \; .
\end{align}

The strip operators are supported on an open surface, and gauge-invariance is ensured by the foliated nature of the theory in the bulk. In other words, we can define the surface operator on a strip $\sigma^i$, generated by a closed curve $\gamma_i(x^i)$ that swipes the area from $x_1^i$ to $x_2^i$. The foliated strip operators are
\begin{align}\label{eq:XYstripsFoliated}
&\Tilde{U}_{\tilde{\beta}}^{i}(x_1^{i},x_2^{i})[\sigma^{i}] = \exp \left\{ i {\tilde{\beta}}\int_{x_1^{i}}^{x_2^{i}} \left( \oint_{\gamma_i(x^i)} b + d B^{i} \right) \right\} \; , \nonumber \\[4pt]
&\Tilde{V}_{\tilde{\alpha}}^{i} (x^{i}_1,x^{i}_2)[\sigma^{i}] = \exp \left\{ i {\tilde{\alpha}} \int_{x^{i}_1}^{x^{i}_2} \left( \oint_{\gamma_i(x^{i})}C^{j} \wedge dx^{j} + d \left(c_j dx^{j}\right) \right) \right\} \; ,
\end{align}
where in $d (c_j dx^{j})$ the index is not summed over: fixing $j$, it means to take the derivative $d$ only of the vector aligned along $dx^j$ with component $c_j$, effectively selecting some directions. To better clarify it, the expression for $i=x$ reads
\begin{align}
    &\Tilde{U}_{\tilde{\beta}}^{x}(x_1,x_2)[\sigma^{x}] = \exp \left\{ i {\tilde{\beta}}\int_{x_1}^{x_2} \left( \oint_{\gamma(x)} b + d B^{x} \right) \right\} \; , \nonumber \\[4pt]
    &\Tilde{V}_{\tilde{\alpha}}^x (x_1,x_2)[\sigma^{x}] = \exp \left\{ i {\tilde{\alpha}} \int_{x_1}^{x_2} \left( \oint_{\gamma(x)}C^{x} \wedge dy + d \left(c_y dy\right) \right) \right\} \; ,
\end{align}
with $d \left(c_y dy\right) = (\partial_x c_y dx + \partial_t c_y dt + \partial_r c_y dr )\wedge dy$. For $i=y$ one only needs to replace $x \leftrightarrow y$.

Their exotic expression has the form
\begin{align}\label{eq:XYstripsExotic}
    \Tilde{V}^{i}_{\tilde{\alpha}}(x^{i}_1,x^{i}_2) [\sigma^{i}] &= \exp \left\{ i \tilde{\alpha} \int_{x^{i}_1}^{x^{i}_2} \left( \oint_{\gamma_i(x^{i})} A_{xy}dx^{j} + \partial_i A_r dr + \partial_i A_t dt \right) dx^{i} \right\} \; , \quad j \neq i \; , \nonumber \\[5pt]
    \Tilde{U}^{i}_{\tilde{\beta}}(x^{i}_1,x^{i}_2) [\sigma^{i}] &= \exp \left\{ i \tilde{\beta} \int_{x^{i}_1}^{x^{i}_2} \left( \oint_{\gamma_i(x^{i})}\Tilde{A}_{xy}dx^{j} + \partial_i \Tilde{A}_r dr + \partial_i \Tilde{A}_t dt \right) dx^{i} \right\} \; , \quad j \neq i \; .
\end{align}
As an example, for $i=x$ they become
\begin{align}
    \Tilde{V}^{x}_{\tilde{\alpha}}(x_1,x_2) [\sigma^{x}] &= \exp \left\{ i \tilde{\alpha} \int_{x_1}^{x_2} dx \left( \oint_{\gamma(x)} A_{xy} dy + \partial_x A_r dr + \partial_x A_t dt \right) \right\} \; , \nonumber \\[5pt]
    \Tilde{U}^{x}_{\tilde{\beta}}(x_1,x_2) [\sigma^{x}] &= \exp \left\{ i \tilde{\beta} \int_{x_1}^{x_2} dx \left( \oint_{\gamma(x)} \Tilde{A}_{xy} dy + \partial_x \Tilde{A}_r dr + \partial_x \Tilde{A}_t dt \right) \right\} \; .
\end{align}
The foliated operators reduce to the exotic ones by the same reasoning described above.

The strips can be topologically deformed while keeping the end manifolds at $x^i_1$ and $x^i_2$ fixed. We can write the braiding between lines and strips from the e.o.m. with the defect insertion as 
\begin{align}\label{eq:XYBraiding}
    &\langle V_{\alpha}[\gamma] , \tilde{U}_{\tilde{\beta}}[\sigma^{i}] \rangle = \exp \left( 2 \pi i \alpha \tilde{\beta} \, \mathrm{Link}_i(\gamma, \gamma^{i}) \right) \; , \nonumber \\[5pt]
    &\langle U_{\beta}[\gamma] , \tilde{V}_{\tilde{\alpha}}[\sigma^{i}] \rangle = \exp \left( 2 \pi i \beta \tilde{\alpha} \, \mathrm{Link}_i(\gamma, \gamma^{i}) \right) \; .
\end{align}

A peculiar feature of the strips is that if $\gamma_i(x^i)$ is defined at some fixed $x$ or $y$, they reduce to a pair of line operators with opposite direction, one at $x_1^i$ and the second one at $x_2^i$. In this case, it is easy to see that $\tilde{V}_{\alpha} = V_{\alpha}[\gamma|_{x_1}]$ - $V_{\alpha}[\gamma|_{x_2}]$ and $\tilde{U}_{\beta} = U_{\beta}[\gamma|_{x_1}]$ - $U_{\beta}[\gamma|_{x_2}]$. For instance, by defining $\gamma_{(y)}$ at fixed $y$, in the exotic theory we get
\begin{align}
    \Tilde{V}^{x}_{\tilde{\alpha}}(x_1,x_2) [\sigma^{x}] &= \exp \left\{ i \tilde{\alpha} \int_{x_1}^{x_2} dx \left( \oint_{\gamma(x)} \partial_x A_r dr + \partial_x A_t dt \right) \right\} \nonumber \\[4pt]
    &= \exp \left\{ i \tilde{\alpha} \left( \oint_{\gamma(x_2)} A_r dr + A_t dt \right) - i \tilde{\alpha} \left( \oint_{\gamma(x_1)} A_r dr + A_t dt \right) \right\} \; ,
\end{align}
and similarly for $\Tilde{U}^{x}$.

\subsubsection*{Boundary operators}

Lines and strips that end or project on the conformal boundary correspond to the operators that generates the dipole symmetries and their charges. In facts, the gapped boundary forces flux quantization as
\begin{align}\label{eq:XYFluxQuantization}
    & \int_{\Sigma \subset M_0} b \in \frac{2 \pi}{R} \mathbb{Z} \; , \nonumber \\[5pt]
    & \int_{x_1^{i}}^{x_2^{i}} \oint_{\gamma \subset M_0} b + dB^{i} \in \frac{2 \pi}{R} \mathbb{Z} \; .
\end{align}
The set of lines $U_{\alpha}$ and $V_{\beta}$ in Eq. \eqref{eq:XYFoliatedLines} trivializes if $\alpha = p /R$ and $ \beta = q^i R$, with $p$, $q^i \in \mathbb{Z}$, thereby ending on the boundary. The non-trivial lines projected on the boundary are identified as 
\begin{align}
    \alpha \sim \alpha + 1/R \; , \quad \beta \sim \beta + R \; ,
\end{align}
and these generate the $U(1) \times U(1)$ dipole symmetries of the XY-plaquette. 

On the other hand, when the strip operator $\tilde{V}$ ends on the boundary, it corresponds to the operators charged under the momentum dipole $U(1)$ symmetry 
\begin{align}
    \exp \left\{ i \int_{x^{i}_1}^{x^{i}_2} \left( \oint_{\gamma_i(x^{i})}C^{i} \wedge dx^{i} + d \left(c_i dx^{i}\right) \right) dx^{i} \right\} = \exp \left\{ i \int_{x^{i}_1}^{x^{i}_2} Q^{i}_m(x^{i}) \right\} \, ,
\end{align}
while $\tilde{U}$ corresponds to operators charged under the winding $U(1)$
\begin{align}
    \exp \left\{ i \int_{x_1^{i}}^{x_2^{i}} \left( \oint_{\gamma_i(x^i)} b + d B^{i} \right) dx^{i} \right\} = \exp \left\{ i \int_{x^{i}_1}^{x^{i}_2} Q^{i}_m(x^{i}) \right\} \, ,
\end{align}
both carrying integer charges \cite{Seiberg:2020bhn}.

\subsection{XYZ-cube model}

Defined on a 3+1 dimensional spacetime $M$, the XYZ-cube model has action \cite{Gorantla:2020xap}
\begin{equation}\label{eq:XYZcube}
    S_{xyz}= \frac{1}{2\pi}\int_M d^4x \frac{\mu_0}{2} \left(\partial_t \phi \right)^2 + \frac{1}{2\mu} \left( \partial_x\partial_y\partial_z \phi \right),
\end{equation}
with $\phi$ a compact scalar field with radius $R$, subject to the identification $\phi\sim\phi+2\pi (n_{xy}(x,y) + n_{yz}(y,z) + n_{xz}(x,z))$. Space rotations are broken down to $S_4$ symmetry, and the scalar field transform in the trivial representation of $S_4$. The continuous subsystem symmetry of this theory are a quadrupole $U(1)$ momentum and a quadrupole $U(1)$ winding symmetry. A dual description can be written in terms of a scalar field $\phi^{xyz}$, in the sign representation of $S_4$, whose action reads
\begin{align}
    S_{xyz} = \frac{1}{2\pi}\int_M d^4x \frac{\tilde{\mu_0}}{2} \left( \partial_t \phi^{xyz} \right)^2 + \frac{1}{2\tilde{\mu}} \left( \partial_x \partial_y \partial_z \phi^{xyz} \right),
\end{align}
where the couplings are expressed as
\begin{align}\label{eq:XYZduality}
    \tilde{\mu}_0 &= \frac{\mu}{4 \pi^2} \; , \nonumber \\[4pt]
    \tilde{\mu} &= 4 \pi^2 \mu_0 \; .
\end{align}
and momentum and winding modes are exchanged. 

The theory can be described in the SymTFT framework with a bulk action
\begin{align}\label{eq:XYZExoticSymTFT}
    S_{exo} = \frac{i}{2\pi} \int_{M\times I} d^5x &\left[ A_t \left( \partial_{r} \tilde{A}_{xyz} -\partial_x \partial_y \partial_z \tilde{A}_r \right) + A_r \left( \partial_x \partial_y \partial_z \tilde{A}_t - \partial_{t} \tilde{A}_{xyz} \right) \right. \nonumber \\[4pt]
    & \left. + A_{xyz} \left( \partial_t \tilde{A}_r - \partial_r \tilde{A}_t \right) \right] \; ,
\end{align}
where $r$ is the extra radial coordinate, and both $A$ and $\Tilde{A}$ are $\mathbb{R}$-gauge field with gauge transformations given by
\begin{align}
    &\delta A_t = \partial_t \lambda \; , \quad \delta A_r = \partial_r \lambda \; , \nonumber \\[4pt]
    &\delta A_{xyz} = \partial_x \partial_y \partial_z \lambda \; , \nonumber \\[4pt]
    &\delta \tilde{A}_t = \partial_t \tilde{\lambda} \; , \quad \delta \tilde{A}_r = \partial_r \tilde{\lambda} \; , \nonumber \\[4pt]
    &\delta \tilde{A}_{xyz} = \partial_x \partial_y \partial_z \tilde{\lambda} \; ,
\end{align}
At the topological boundary, a compact scalar edge mode is coupled to the gauge fields as
\begin{align}\label{eq:XYZTopologicalBoundary}
    S_0 = \frac{iR}{2\pi} \int_{M_0} d^4x \, \phi \left(\partial_x \partial_y \partial_z \tilde{A}_t - \partial_{t} \tilde{A}_{xyz} \right) \; ,
\end{align}
with $\delta \phi=-\frac{\lambda}{R}$, and the b.c. read
\begin{align}\label{eq:XYZTopologicalBC}
    A_t &= R \, \partial_t \phi \nonumber \\[4pt]
    A_{xyz}&= R \, \partial_x \partial_y \partial_z \phi \; .
\end{align}
On the conformal boundary at $r=L$ we have 
\begin{align}\label{eq:XYZConformalBoundary}
    S_L = \frac{1}{4\pi} \int_{M_L} d^4x \, g {\tilde{A}_t}^2 + {\tilde{A}_{xyz}}^2 \; ,
\end{align}
enforcing the b.c.
\begin{align}\label{eq:XYZconformalBC}
     &\tilde{A}_{xyz} = i A_t   \; , \nonumber \\[4pt]
     &\tilde{A}_t = - \frac{i}{g} A_{xyz} \; .
\end{align}
Upon slab compactification and imposing the above b.c. we get the XYZ mode in eq. \eqref{eq:XYZcube} with $\mu_0 = R^2$ and $1/ \mu = R^2/g$.

As for the foliated SymTFT description, the bulk involves single and double foliations, with action 
\begin{align}
    S_{fol} = \frac{i}{2\pi} & \int_{M \times I} \sum_i \left(C^i\wedge dB^i\wedge dx^i + b\wedge C^i\wedge dx^i \right) \nonumber \\[4pt]
    & + \sum_{ij} \left( \mathbf{C}^{ij} \wedge d\mathbf{B}^{ij} \wedge dx^i\wedge dx^j + B^i\wedge\mathbf{C}^{ij} \wedge dx^i \wedge dx^j \right) + c\wedge db \; ,
\end{align}
where b is a 3-form, the $B^i$ are 2-forms, and $\mathbf{B}^{ij}$, $\mathbf{C}^{ij}$, $C^i$ and $c$ are 1-forms, with all field real valued. By integrating out the $r$ components of $b$, $C^i$ and $\mathbf{C}^{ij}$, this action reduces to eq. \eqref{eq:XYZExoticSymTFT}. The gauge transformations are
\begin{equation}
    \begin{split}
        \delta b&=d\chi_2 \; , \\
        \delta B^i&=d\chi_1^i-\chi^2 \; ,\\
        \delta\mathbf{B}^{ij}&=d\chi_0^{ij}-\chi_1^i-\chi_1^j \; ,\\
        \delta c&=d\lambda-\sum_i\lambda^i\wedge dx^i \; ,\\
        \delta C^i&=d\lambda^i-\sum_j\lambda^{ij}\wedge dx^jj \; ,\\
        \delta \mathbf{C}^{ij}&=d\lambda^{ij} \; .
    \end{split}
\end{equation}
The gapped boundary has form
\begin{align}\label{eq__xyzgapbdy}
    S_0 =\frac{iR}{2\pi} & \int_{M_0} \sum_{i} \left( \Phi^i \wedge dB^i \wedge dx^i + \Phi^i \wedge b\wedge dx^i \right)\\
    & + \sum_{i,j} \left( \mathbf{\Phi}^{ij} \wedge d\mathbf{B}^{ij} \wedge dx^i\wedge dx^j + \mathbf{\Phi}^{ij} \wedge B^i \wedge dx^i \wedge dx^j \right) + \phi\wedge db \; , 
\end{align}
where $\phi$, $\Phi^i$ and $\mathbf{\Phi}^{ij}$ are compact scalars of radius $R$ with gauge transformations $\delta \phi=-\frac{\lambda}{R}$, $\delta \Phi^i=-\frac{\lambda^i}{R}$, and $\delta \mathbf{\Phi}^{ij}=-\frac{\lambda^{ij}}{R}$. This boundary yields the conditions
\begin{align}\label{eq:XYZFoliatedTopBC}
        c&=  - R \left( d\phi+\Phi^i\wedge dx^i \right) \; , \nonumber \\[4pt]
        C^i\wedge dx^i&= R \left( d\Phi^i \wedge dx^i - \mathbf{\Phi}^{ij} \wedge dx^i \wedge dx^j \right) \; , \nonumber \\[4pt]
        \mathbf{C}^{ij} \wedge dx^i \wedge dx^j&= - R \, d\mathbf{\Phi}^{ij} \wedge dx^i \wedge dx^j \; .
\end{align}
Finally, on the conformal boundary we have 
\begin{align}\label{eq__xyzscalebdy}
    S_L = \frac{1}{4\pi} \int_{M_L} & \left\{ b+\sum_{ijk}d(B^i_{jk}dx^j\wedge dx^k) + \sum_{ijk} d\left[ d (\mathbf{B}^{ij}_kdx^k)_{jk}dx^j \wedge dx^k \right] \right\} \wedge *_3 \{ b + \ldots \} \wedge dt \nonumber \\[4pt]
    &+(\mathbf{B}^{xy}-\mathbf{B}^{yz}+\mathbf{B}^{zx}) \wedge *_1 \left( \mathbf{B}^{xy}-\mathbf{B}^{yz} + \mathbf{B}^{zx} \right) \wedge dx \wedge dy \wedge dz \; .
\end{align}
The conformal b.c. reads
\begin{align}\label{eq:XYZconformalBoundary}
    &c = i *_3 \left\{ \ldots \right\} \; , \nonumber \\[4pt]
    &C_t^x = + i \partial_x \left\{ \ldots \right\} \; , \quad C_t^y = - i \partial_y \left\{ \ldots \right\} \; , \quad C_t^z = + i \partial_z \left\{ \ldots \right\} \; , \nonumber \\[4pt]
    &\mathbf{C}_t^{xy} = + i \partial_x \partial_y \left\{ \ldots \right\} \; , \quad \mathbf{C}_t^{xz} = - i \partial_x \partial_z \left\{ \ldots \right\} \; , \quad \mathbf{C}_t^{y z} = + i \partial_y \partial_z \left\{ \ldots \right\} \; , \nonumber \\[4pt]
    &\mathbf{C}_t^{xy} = - i \left( \mathbf{B}_t^{xy} - \mathbf{B}_t^{yz} + \mathbf{B}_t^{zx} \right) \; , \nonumber \\[4pt]
    &\mathbf{C}_t^{yz} = + i \left( \mathbf{B}_t^{xy} - \mathbf{B}_t^{yz} + \mathbf{B}_t^{zx} \right) \; , \nonumber \\[4pt]
    &\mathbf{C}_t^{zx} = + i \left( \mathbf{B}_t^{xy} - \mathbf{B}_t^{yz} + \mathbf{B}_t^{zx} \right) \; , \nonumber \\[4pt]
    & C^x_y = C^x_z = C^y_x = C^y_z = C^z_x = C^z_y = 0 \; ,
\end{align}
where 
\begin{equation}
    \left\{ \ldots \right\} = \left\{ b+\sum_{ijk}d(B^i_{jk}dx^j\wedge dx^k) + \sum_{ijk} d\left[ d (\mathbf{B}^{ij}_kdx^k)_{jk}dx^j \wedge dx^k \right] \right\} \; .
\end{equation}
After compactification, integrating out the scalars $\Phi^i$ and $\mathbf{\Phi}^{ij}$ we obtain the XYZ-cube theory \eqref{eq:XYZcube}, whereas by integrating out $\phi$, setting $b=da$, we obtain a foliated dual of the XYZ-cube
\begin{align}\label{eq:XYZFoliatedDual}
    S =  \frac{1}{4\pi} \int_{M} & d\left( a + \sum_{ijk} B^i_{jk}dx^j\wedge dx^k + \sum_{ijk} \left( \mathbf{B}^{ij}_kdx^k\right)_{jk} dx^j  \wedge dx^k \right) \wedge *_3 d( a + \ldots) \wedge dt \nonumber \\[5pt]
    &+(\mathbf{B}^{xy}-\mathbf{B}^{yz}+\mathbf{B}^{zx}) \wedge *_1 \left( \mathbf{B}^{xy}-\mathbf{B}^{yz} + \mathbf{B}^{zx} \right) \wedge dx \wedge dy \wedge dz \; , \nonumber \\[5pt]
    & \Phi^i\wedge d(a + B) \wedge dx^i + \mathbf{\Phi}^{ij} \wedge \left( B^i + d\mathbf{B}^{ij} \right) \wedge dx^i\wedge dx^j
\end{align}
which can be interpreted as a foliated version of the 2-form Maxwell gauge theory in 4d with gauge field $( a + \sum_{ijk} B^i_{jk}dx^j\wedge dx^k + \sum_{ijk} \left( \mathbf{B}^{ij}_kdx^k\right)_{jk} dx^j  \wedge dx^k )$, coupled to single and double foliated BF theory.

\subsubsection*{Partially topological operators}

The set of (partially) topological operators consists of line and bar operators. The lines are supported on a closed $\gamma$ at fixed $(x,y,z)$ as 
\begin{align}\label{eq:XYZLines}
     V_{\alpha}(x,y,z)[\gamma]&= \exp \left( i\alpha \oint A_tdt + A_r dr \right) \nonumber \\[4pt]
     & = \exp \left( i\alpha\oint_\gamma c\right) \nonumber \\[5pt]
     U_{\beta}(x,y,z)[\gamma]&= \exp \left( i\beta\oint \tilde{A}_t dt + \tilde{A}_r dr \right) \; , \nonumber \\[4pt]
     &=\exp \left( i\beta\oint_\gamma B^{xy} -B^{yz} + B^{zx} \right) \; .
\end{align}
The bar operators $\rho^{ij}$ are defined as a line $\gamma$ that swipes the intervals $[x_1^i, \, x_2^i]$ and $[x_1^j, \, x_2^j]$, written as
\begin{align}\label{eq:XYZbars}
    \Tilde{V}_{\alpha}&(x^i_1,x^i_2,x^j_1,x^j_2)[\rho^{ij}]=\exp\left(i\alpha\int^{x^i_2}_{x^i_1} dx^i \int^{x^j_2}_{x^j_1} dx^j \oint_{\gamma(x^i,x^j)} \partial_i\partial_j A_tdt + A_{xyz}dx^{k}\right) \nonumber \\[4pt]
    &= \exp \left( i \int^{x^i_2}_{x^i_1} \int^{x^j_2}_{x^j_1} \oint C^{ij}\wedge dx^i\wedge dx^j+d[C^i\wedge dx^i + (d(c_idx^i))_{ij}dx^i\wedge dx^j] \right) \; , \nonumber \\[4pt]  
    \Tilde{U}_{\beta}&( x^i_1,x^i_2,x^j_1,x^j_2)[\rho^{ij}]=\exp \left( i\beta\int^{x^i_2}_{x^i_1} dx^i \int^{x^j_2}_{x^j_1} dx^j \oint_{\gamma(x^i,x^j)}\partial_i\partial_j \tilde{A}_tdt + \tilde{A}_{xyz}dx^{k} \right) \nonumber \\[4pt]
    &= \exp \left( i \beta \int^{x^i_2}_{x^i_1} \int^{x^j_2}_{x^j_1} \oint b + \sum_{mnk} d(B^m_{nk}dx^n \wedge dx^k) + \sum_{mnk} d(d(\mathbf{B}^{mn}_kdx^k)_{nk}dx^n \wedge dx^k) \right) \; ,
\end{align}
where in $d(d(\mathbf{B}^{ij}_kdx^k)_{jk}$ means that we are first taking derivative of the vector aligned along $dx^k$, then performing a second derivative of tensor whose only components are in $dx^j \wedge dx^k$. Similarly for the terms $d[(d(c_idx^i))_{ij}dx^i\wedge dx^j]$ and $d(B^i_{jk}dx^j \wedge dx^k)$. As an example, take $x^i=x$ and $x^j=y$, the $U$ operator is
\begin{equation}
    \begin{split}
         \Tilde{U}_{\beta}&( x_1,x_2,y_1,y_2)[\rho^{xy}]=\exp \left( i\beta\int^{x_2}_{x_1} dx \int^{y_2}_{y_1} dy \oint_{\gamma(x,y)}\partial_z\partial_y \tilde{A}_tdt + \tilde{A}_{xyz}dz \right) \nonumber \\[4pt]
    &= \exp \left( i \beta \int^{x_2}_{x_1}\int^{y_2}_{y_1}\oint_{\gamma(x,y)} b + \sum_{mnk} d(B^m_{nk}dx^n \wedge dx^k) + \sum_{mnk} d(d(\mathbf{B}^{mn}_kdx^k)_{nk}dx^n \wedge dx^k) \right)  .
    \end{split}
\end{equation}
To further clarify the notation on the foliated case, we write the operator in the case where $\gamma(x,y)$ is a close loop in the $z$ direction:
\begin{align}
         \Tilde{U}_{\beta}(x_1,x_2,y_1,y_2)[\rho^{xy}]= \exp & \left( i \beta \int^{x_2}_{x_1}dx\int^{y_2}_{y_1}dy\oint dz \;   b_{xyz}+\partial_{x}B^x_{yz}-\partial_{z}B^r_{xy}+\partial_{y}B^y_{zx} \right. \nonumber \\[4pt]
    & \left. + \partial_x\partial_y\mathbf{B}^{xy}_z-\partial_x\partial_z\mathbf{B}^{xz}_y+\partial_y\partial_z\mathbf{B}^{yz}_x \right) \; .
    \end{align}

Their linking reads
\begin{equation}\label{eq::3+1quadripolelink}
    \begin{split}
        \langle \Tilde{V}_{\alpha}( x^i_1,x^i_2,x^j_1,x^j_2)[\rho^{ij}],U_{\beta}(x,y,z)[\gamma] \rangle&=e^{2\pi i\alpha\beta \rm Link_{ij}(\gamma,\rho^{ij})} \; , \\[4pt]
        \langle \Tilde{U}_{\alpha}( x^i_1,x^i_2,x^j_1,x^j_2)[\rho^{ij}],V_{\beta}(x,y,z)[\gamma]\rangle&=e^{2\pi i\alpha\beta {\rm Link}_{ij}(\gamma,\rho^{ij})} \; .
    \end{split}
\end{equation}

\subsubsection*{Boundary operators} 

The topological boundary conditions eq. \eqref{eq:XYZTopologicalBC} impose flux quantization 
\begin{align}
    &\oint b \in\frac{2\pi}{R}\mathbb{Z} \; , \nonumber\\[4pt]
    &\int \oint B^i\wedge dx^i \in\frac{2\pi}{R}\mathbb{Z} \; , \nonumber \\[4pt]
    &\int \int \oint \mathbf{B}^{ij}\wedge dx^i\wedge dx^j\in\frac{2\pi}{R}\mathbb{Z} \; . 
\end{align}
Lines $V_{\alpha}$ are trivialized on the boundary if $\alpha = 2 \pi R n$, while $U_{\beta}$ with $\beta = 2 \pi m /R$, $n, \, m \in \mathbb{Z}$. The non trivial lines are identified as $\alpha \sim \alpha + R$, $\beta \sim \beta + 1/R$ and they generate the quadrupole current operators. The bar operators become the charges under momentum quadrupole
\begin{align}
    \Tilde{V}_\alpha(x^i_1,x^i_2,x^j_1,x^j_2)=\exp \left( i\alpha\int^{x^i_2}_{x^i_1}\int^{x^j_2}_{x^j_1}Q^{ij} \right) \; ,
\end{align}
and under winding quadrupole
\begin{align}
    \Tilde{U}_\beta(x^i_1,x^i_2,x^j_1,x^j_2)=\exp \left( i \beta \int^{x^i_2}_{x^i_1}\int^{x^j_2}_{x^j_1}Q^{xyz}_k \right) \; .
\end{align}

\section{Duality of the XY plaquette model}\label{sec:GenDuality}

In this section we discuss the presence of an exotic $\theta$-term for the XY-plaquette model, using the SymTFT \textit{Mille-feuille} of the previous section to add this term to the topological boundary. We proceed by generalizing the self-duality of \cite{Seiberg:2020bhn} in presence of such term. Combining this self-duality with gauging, a continuous duality symmetry is generated. 

\subsection{Exotic theta term}\label{sec:exotictheta}

The gapped boundary of the 3+1 SymTFT in eq. \eqref{eq:XYTopologicalBoundary} admits a discrete torsion term, which modifies the topological boundary conditions. Upon slab compactification, the effect is to add an exotic $\theta$-term to the XY-plaquette model. Let us write the gapped boundary as follow
\begin{equation}\label{eq:XYThetaTopologicalBoundary}
\begin{split}
    S_{0} = \int_{M_0} d^3x \, \frac{iR}{2\pi} \phi \left( \partial_t \Tilde{A}^{xy} - \partial_x \partial_y \Tilde{A}^t \right) - \frac{i \theta^{xy}}{4 \pi^2 R} \phi \left( \partial_t A^{xy} - \partial_x \partial_y A^t \right) + \frac{i \theta^{xy}}{4\pi^2 R^2} A_t A_{xy} \; ,
\end{split}
\end{equation}
where $\theta^{xy}$ transforms in the spin-2 representation of the spatial $\mathbb{Z}_4$ rotational symmetry. 
The boundary conditions now read
\begin{align}
    &A_{xy} = - R \partial_{x} \partial_y \phi \; , \quad \tilde{A}_{xy} = - \frac{\theta^{xy}}{R} \partial_{x} \partial_y \phi + \frac{\theta^{xy}}{R^2} A_{xy} \; , \nonumber \\[4pt]
    &A_{t} = - R \partial_t \phi \; , \quad \tilde{A}_{t} = - \frac{\theta^{xy}}{R} \partial_t \phi + \frac{\theta^{xy}}{R^2} A_{xy} \; ,
\end{align}
and they respect the flux quantization condition in eq. \eqref{eq:XYFluxQuantization}. Notice that this boundary conditions ensure that the last term in $S_0$ gives an integer, because of flux quantization of $\oint A$ they force. In order to see this, we fix the gauge so that $A_t$ only depends on $t$ and $A_{xy}$ only depends on $x$ and $y$,
\begin{align}
    &\partial_x \left( \partial_t \lambda \right) = - \partial_x A_t \; , \nonumber \\
    &\partial_y \left( \partial_t \lambda \right) = - \partial_y A_t \; , \nonumber \\
    &\partial_t \left( \partial_x \partial_y \lambda \right) = - \partial_t A_{xy} \; .
\end{align}
With this choice
\begin{equation}
    \frac{i\theta^{xy}}{2\pi R^2} \int_{M_L} dtdxdy A_t A_{xy} = \frac{i\theta^{xy}}{2\pi R^2} \oint dt A_t \oint dx dy A_{xy} \in 2\pi \theta^{xy} \mathbb{Z} \; ,
\end{equation}
therefore, the theta term also has a periodic identification $\theta^{xy}\sim\theta^{xy}+2\pi n_{x}(x)-2\pi n_{y}(y)$.
The bulk of the SymTFT and the physical boundary are not modified. After compactification, where we put $g=1$ for simplicity, we obtain the XY-plaquette with an exotic $\theta$ term
\begin{equation}\label{eq::xylagrangian_theta}
    S = \int_{M} d^3x \, \frac{R^2}{4\pi} \left( \partial_t \phi \right)^2 + \frac{R^2}{4 \pi} \left( \partial_x \partial_y \phi \right)^2 + \frac{i \theta^{xy}}{4 \pi^2} \, \partial_t \phi \, \partial_x\partial_y \phi \; .
\end{equation}
Interestingly, such term can be turned on due to the presence of an even number of spatial derivatives in the action. For this reason, this term is not allowed in the XYZ-cube model.

As we mentioned above, in \cite{Seiberg:2020bhn} they show that the XY-plaquette enjoys a self-duality that sends the coupling $(\mu_0, \mu) \to (\tilde{\mu}_0 , \tilde{\mu}) = (\mu/(4 \pi^2), 4 \pi^2 \mu_0)$, which given the identification $\mu_0 = R^2$ corresponds to the inversion of the radius $R$ as in T-duality. This duality can be generalized in presence of the exotic $\theta$-term. In order to see that, we can rewrite the action in eq. \eqref{eq::xylagrangian_theta} in a compact way by introducing the field $(D^{+} \phi, D^{-} \phi)$, with $D^{\pm} \phi = \partial_t \phi \pm \partial_x \partial_y \phi $. Notice that $D^{\pm} \phi$ transform as a composition of the one-dimensional representations of $\mathbb{Z}_4$ trivial and sign, i.e. $\mathbf{1}_0 \pm \mathbf{1}_2$. Hence, the field $(D^{+} \phi, D^{-} \phi)$ transforms as a reducible two-dimensional representation $(\mathbf{1}_0 + \mathbf{1}_2, \, \mathbf{1}_0 - \mathbf{1}_2)$, whose components are exchanged under a $\mathbb{Z}_4$ transformation\footnote{From the SymTFT bulk description, the same choice holds with the gauge fields $A^{\pm} = A_t \pm A_{xy}$. These two fields are exchanged by the discrete rotational $\mathbb{Z}_4$ symmetry.}. In the same spirit, we introduce a complex coupling $(\bar{\tau}, - \tau)$ with $\tau = \frac{\theta^{xy}}{2 \pi} + i R^2 $, where the two components are exchanged under a discrete spatial rotation. The action of the XY-plaquette can be rewritten as
\begin{align}\label{eq:XYCompact}
    S = \frac{i}{8\pi} \int_M d^3x \; \bar{\tau} (D^+ \phi)^2 - \tau(D^- \phi)^2 \; .
\end{align}

In the compactified theory, we can perform the gauging of the shift symmetry by introducing a gauge field in the same representation $(\chi^+,\chi^-)$, coupled to the dual field $\phi^{xy}$ as 
\begin{align}
    S =  \int_M d^3x \, \frac{1}{8\pi} \bar{\tau} (D^+ \phi - \chi^+)^2 - \frac{1}{8\pi} \tau(D^- \phi - \chi^-)^2 - \frac{i}{4\pi} D^+ \phi^{xy} \chi^+ + \frac{i}{4\pi} D^- \phi^{xy} \chi^-  \; .
\end{align}
Since $\phi^{xy}$ is in the sign representation of $\mathbb{Z}_4$, the doublet dual field is $(D^{+}\phi^{xy}, \, - D^- \phi^{xy})$, namely the terms with time derivative transform as $\mathbf{1}_2$ of $\mathbb{Z}_4$. This is why $-D^- \phi^{xy}$ is coupled with $\chi^-$. By integrating out the gauge fields $\chi^{\pm}$, we get
\begin{align}\label{eq:XYCompactDual}
    S = \frac{i}{8\pi} \int_M d^3x \, -\frac{1}{\bar{\tau}} (D^+ \phi^{xy})^2 + \frac{1}{\tau}(D^- \phi^{xy})^2 \; ,
\end{align}
with the condition $\partial_t \partial_x \partial_y \phi^{xy} = 0$. The action in eq. \eqref{eq:XYCompactDual} in terms of the dual field has coupling $(-1/\bar{\tau}, 1/\tau)$, with the two components exchanged under a spatial rotation. This generalizes the duality of \cite{Seiberg:2020bhn}. Interestingly, the XY-plaquette shows a duality similar to Maxwell, while its foliated description derived in eq. \eqref{eq:FoliatedMaxwell} has the form of a foliated Maxwell theory.

\subsection{Gauging and duality symmetry}

We perform a gauging procedure in two steps to show that the XY-plaquette model, with the $\theta$-term turned on, enjoys a continuous duality symmetry. Let us start with the compact action in eq. \eqref{eq:XYCompact}, with coupling $\tau=R^2 + i \frac{\theta^{xy}}{2 \pi}$, and gauge the entire winding symmetry by coupling the current $(D^+ \phi, \, - D^- \phi)$ with a gauge field $\eta = (\eta^+ , \, \eta^- )$ valued in $U(1)$ as
\begin{align}
    S &= \frac{i}{8\pi} \int_M d^3x \; \bar{\tau} (D^+ \phi)^2 - \tau(D^- \phi)^2 + \frac{i}{4 \pi} \int_M d^3x (D^+ \phi) \eta^+ - (D^- \phi) \eta^- \nonumber \\[4pt]
    &- \frac{i}{4 \pi R^2} \int_M d^3x (D^+ b) \eta^+ - (D^- b) \eta^- \; ,
\end{align}
where $b \in \mathbb{R}$ ensures that $D^+ \eta^- = D^- \eta^+ = 0$, morally a flat condition on $\eta$. By integrating out the gauge field $\eta$, we get
\begin{align}
    S = \frac{1}{8\pi} \int_M d^3x \; (D^+ b)^2 + (D^- b)^2 = \frac{1}{4\pi} \int_M d^3x \; (\partial_t b)^2 + (\partial_{x} \partial_y b)^2 \; ,
\end{align}
where the term with $\theta^{xy}$ vanishes for $b$ being non-compact. 

As a second step, we introduce a real-valued field $c = (c^+, c^-)$ to gauge a subgroup $\mathbb{Z}$ of the shift symmetry. We couple $c$ with a compact field $\tilde{\phi} \in U(1)$ in order to impose discrete fluxes on $c$. In particular, the field $\tilde{\phi}$ is chosen to have the same properties of the original $\phi$ field, i.e. $\tilde{\phi} \sim \tilde{\phi} + 2 \pi \left( \tilde{n}_x(x) +  \tilde{n}_y(y)\right)$ with period $\tilde{R}$. We write 
\begin{align}
    S &= \frac{1}{8\pi} \int_M d^3x \; (D^+ b - c^+)^2 + (D^- b - c^-)^2 - \frac{i}{4\pi} \tilde{R}^2 \int_M d^3x \; c^+ \left( D^+ \tilde{\phi} \right) + c^- \left( D^- \tilde{\phi} \right) \nonumber \\[4pt]
    &+ \frac{i \tilde{\theta}^{xy}}{16\pi^2} \tilde{R}^2 \int_M d^3x \; \left( D^+ \tilde{\phi} \right) ^2 - \left( D^- \tilde{\phi} \right)^2 \; ,
\end{align}
where we added a discrete torsion term for $\tilde{\phi}$. Integrating over the field $c$ yields
\begin{align}
    S = \frac{i}{8\pi} \int_M d^3x \; \left( \frac{\tilde{\theta}^{xy}}{2 \pi} - i \tilde{R}^2 \right) \left( D^+ \tilde{\phi} \right)^2 - \left( \frac{\tilde{\theta}^{xy}}{2 \pi} + i \tilde{R}^2 \right) \left( D^+ \tilde{\phi} \right)^2 \; ,
\end{align}
which is the action of a $XY$-plaquette model with coupling $\tilde{\tau} = \frac{\tilde{\theta}^{xy}}{2 \pi} + i \tilde{R}^2 $. We have the freedom to choose the value of this coupling, in particular, after performing the duality transformation, we can take the coupling back to the the original value $\tau$, for a full circle that defines a duality symmetry. In the next section, we construct the condensation defects in the SymTFT that generates this duality symmetry, and show that it has non-invertible fusion rules.

\section{Condensation defects} \label{sec:CondDefects}

In this section we characterize the condensation defects that generate the 0-form symmetries of the bulk theories, relating them to the symmetries of the XY plaquette and XYZ cube. In sec. \ref{sec::symtft}, we reviewed the bulk descriptions of these models, i.e. the dual foliated and exotic. In order to implement higher-gauging of $\mathbb{R} \times \mathbb{R}$ on a codim-1 surface of the bulk $\Sigma$, one condenses these topological operators on $\Sigma$ and sum on their possible configurations. However, it is not clear how to properly define condensation of the topological operators in the foliated theory, as these are partially topologically and $\Sigma$ extend along the directions no longer Lorentz-invariant, i.e. $(x,y)$. To be more precise, let us focus on the 3+1d SymTFT, whose topological operators are lines eq. \eqref{eq:XYFoliatedLines} and strips eq. \eqref{eq:XYstripsFoliated}. Let be $\Sigma = T^3$  with coordinates $(t,x,y)$. The line operators $V_{\alpha} (x,y) [\gamma]$, $U_{\beta} (x,y) [\gamma]$ can be put along the one dimensional submanifolds $S_1^t$ with fixed $x$ and $y$. Condensing an arbitrary number $N$ of lines on $\Sigma$ requires integrating over all possible insertion points $dxdy$ and summing over the 1-cycles in $S_1^t$. Therefore, in order to condense a line operator $W(x,y)[\gamma]$ we need to consistently define a measure in the space of all configurations, for example
\begin{align}\label{eq:MeasureCondLines}
    \sum_{N=1}^{\infty} \prod_{i=1}^N \int dx_i dy_i \sum_{\gamma \in H_1(S^1, \mathbb{R})} W(x_i, y_i)[\gamma] \;= \sum_{N=1}^{\infty} \prod_{i=1}^N \int dx_i dy_i \int_{\mathbb{R}}d\alpha\; W(x_i, y_i)_\alpha[S_1^t] ,
\end{align}
where $W(x,y)[\gamma]$ can be a composition of $U(x,y)[\gamma]$ and $V(x,y)[\gamma]$ depending on the higher gauging we perform.

\begin{figure}
    \centering
    \includegraphics[scale=0.3, trim={20cm 5cm 20cm 7cm}, clip]{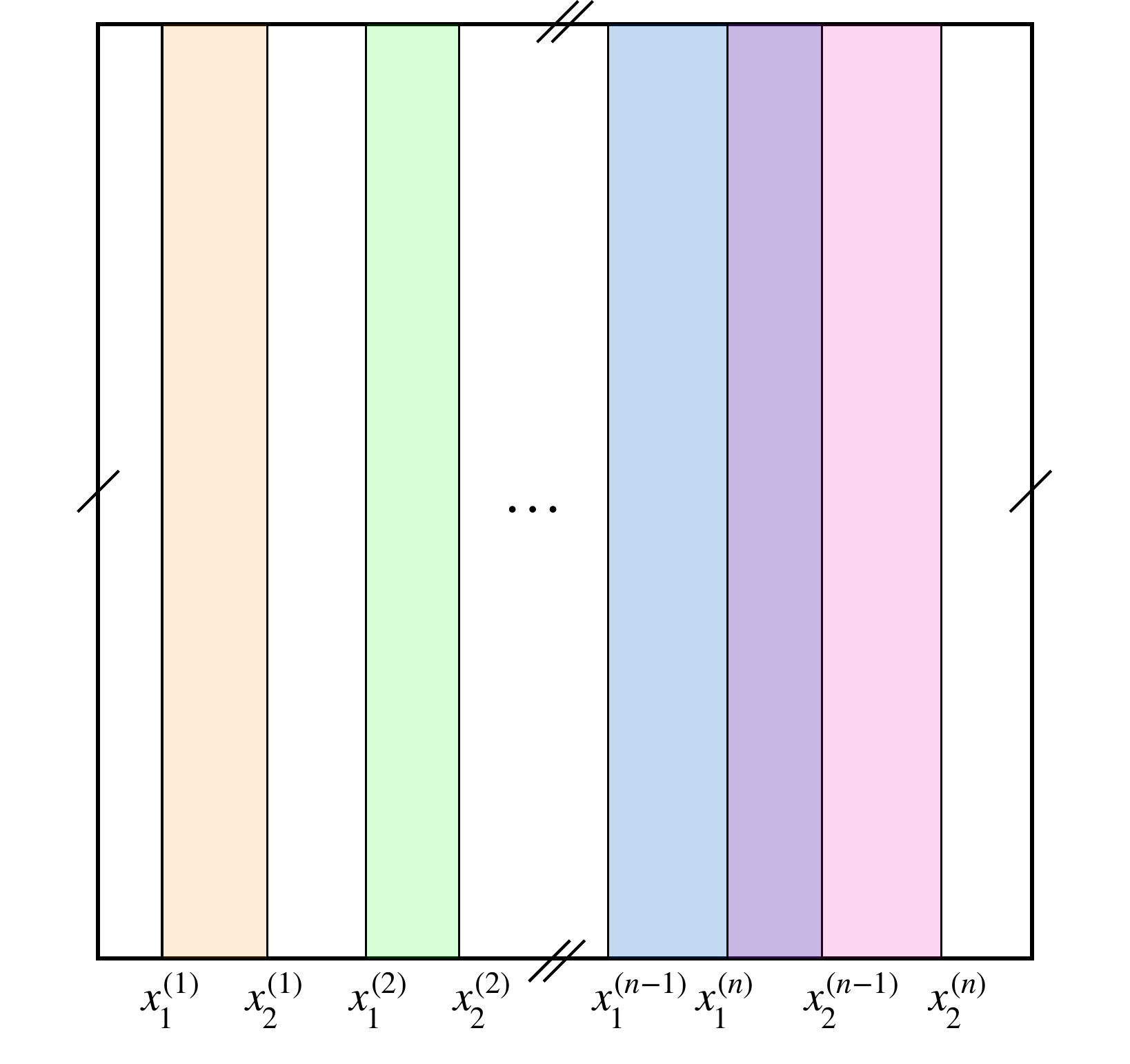}
    \caption{On the condensation defect supported on $T^3$ parametrized by $(x,y,t)$, an arbitrary number $n$ of strips can be inserted at arbitrary positions of a torus $T^2 \subset T^3$, defined at fixed $t$. Some of the strips can overlap. The interval $x$ is divided into an arbitrary set of subintervals.}
    \label{fig:StripsT2}
\end{figure}

The strips are defined as closed lines that swipe the area between two fixed locations at $x_1^i$ and $x_2^i$. When the closed lines are defined along $t$, then the strip is equivalent to the two lines $\gamma(x_1^i)$ and $\gamma(x_2^i)$, the latter with opposite orientation. In this case, the problem of defining the condensation reduces to eq. \eqref{eq:MeasureCondLines}. In any other strip configuration, the operators are defined on a strip of $T^2 \subset T^3$, hence condensing an arbitrary number of such operators splits the $x$ coordinate of the $T^2$ in numerous sub-intervals, see for example Fig. \ref{fig:StripsT2}. To sum over all possible insertions requires defining a measure in the space of all partition of an interval.

Furthermore, the bulk 0-form symmetry of the exotic description is $SL(2,\mathbb{R})$, as we show in the next sub-section. We expect the foliated bulk description to enjoy either the same group symmetry or a larger one containing $SL(2,\mathbb{R})$. However, a direct rotation of the foliated gauge fields breaks the foliation. To see this, notice that a rotation of the one-forms needs to mix up $B^x$, $B^y$ and $c$, but $c$ has components in all directions, while $B^x_x$ and $B^y_y$ are suppressed by the foliation.

Highlighting these open technical questions for condensation defects in foliated theories, we rely only on an explicit lagrangian description for the exotic models. The exotic/foliated duality ensures that the condensation defects exist also for the foliated theories.

In the following, we construct these condensation defects in the bulk, supported on a manifold with boundaries. In deriving the fusion rules of the boundaries, we focus on a subset of the condensation defect, $\mathcal{K}$, that leaves physical boundary conditions invariant and show that the rules are indeed non-invertible. Upon gauging the subgroup of the 0-form symmetries in the bulk generated by $\mathcal{K}$, the condensation defect becomes transparent, while its boundary becomes a genuine operator on the physical theory which takes the role of duality defect \cite{Burbano:2021loy, Kaidi:2022cpf, Antinucci:2022vyk}.

\subsection{Condensation defects in 3+1d and XY plaquette}

In section \ref{sec:XYplaquette} we showed that the SymTFT description of the XY plaquette is given by the 2-foliated theory and the dual exotic theory.

The (3+1)-dimensional bulk theory has lagrangian
\begin{equation}
    \cL=\frac{i}{2\pi}\left[ A^t \left( \partial_r \Tilde{A}_{xy}-\partial_x \partial_y \Tilde{A}_r \right) + A_r \left( \partial_x \partial_y \Tilde{A}_t -  \partial_t \Tilde{A}_{xy} \right) + A_{xy} \left( \partial_r \Tilde{A}_t - \partial_t \Tilde{A}_r \right) \right] \; ,
\end{equation}
whose 0-form symmetry transformation can be obtained by
\begin{equation}\label{eq::sym-action-bulk-xy}
    \begin{pmatrix}
    a & b^{xy} \\
    c^{xy} & d
    \end{pmatrix}
    \begin{pmatrix}
    A'\\
    \Tilde{A}'
    \end{pmatrix}=
    \begin{pmatrix}
    A\\
    \Tilde{A}
    \end{pmatrix} \; ,
\end{equation}
where $b^{xy}$, $c^{xy}$ transform under the sign representation $\mathbf{1}_2$ of the $\mathbb{Z}_4$ spatial rotations, in order to have a meaningful combination of the gauge fields. From now on, for notational and computational simplicity we decompose them as $b \,  \mathbb{1}^{xy}$, where $\mathbb{1}^{xy}$ is a unitary constant that rotates to -1 under a $\mathbb{Z}_4$ rotation. The Lagrangian becomes
\begin{equation}
\begin{split}
    \cL=&\frac{i(ad-bc)}{2\pi}\left[ A^t \left( \partial_r \Tilde{A}_{xy} -\partial_x \partial_y \Tilde{A}_r \right) + A_r \left( \partial_x \partial_y \Tilde{A}_t -\partial_t \Tilde{A}_{xy} \right) + A_{xy} \left( \partial_r \Tilde{A}_t - \partial_t \Tilde{A}_r \right) \right] \nonumber \\[4pt]
    & + \frac{i(ab)\mathbb{1}^{xy}}{2\pi} \left[ A_t \left( \partial_r A_{xy} - \partial_x \partial_y  A_r \right) + A_r \left( \partial_x \partial_y A_t - \partial_t A_{xy} \right) + A_{xy} \left( \partial_r A_t - \partial_t A_r \right) \right] \nonumber \\[4pt]
    &+ \frac{i(cd)\mathbb{1}^{xy}}{2\pi} \left[ \Tilde{A}_t \left( \partial_r  \Tilde{A}_{xy} - \partial_x \partial_y \Tilde{A}_r \right) + \Tilde{A}_r \left( \partial_x \partial_y \Tilde{A}_t - \partial_t \Tilde{A}_{xy} \right) + \Tilde{A}_{xy} \left( \partial_r \Tilde{A}_t - \partial_t \Tilde{A}_r \right) \right] \; .
\end{split}
\end{equation}
It is easy to see that the terms of the second and third lines sum to zero. Therefore, in order to have a symmetry the only constraint is $ad-bc=1$, i.e. the bulk SymTFT has a $SL(2,\R)$ symmetry. By the Iwasawa decomposition, this is generated by the three matrices
\begin{align}\label{eq:Iwasawa}
&\mathcal{K}_{\omega} =
    \begin{pmatrix}
         \cos \omega & - \mathbb{1}^{xy} \, \sin \omega \\
         \mathbb{1}^{xy} \, \sin \omega & \cos \omega \\
    \end{pmatrix} \; , \qquad \omega \in \mathbb{R} \; , \\[4pt]
&\mathcal{T}_{l} =
    \begin{pmatrix}
         1 & 0 \\
         l^{xy} & 1 \\
    \end{pmatrix} \; , \qquad l^{xy} = l \, \mathbb{1}^{xy} \; , \quad l \in \mathbb{R}/\{0\} \; , \\[4pt]
&\mathcal{A}_{s} =
    \begin{pmatrix}
         s & 0 \\
         0 & \frac{1}{s} \\
    \end{pmatrix} \; , \qquad s \in \mathbb{R} \; .
\end{align}
acting on the doublet $(A, \tilde{A})$ of gauge fields.

The conformal boundary conditions \eqref{eq:XYconformalBoundary} are preserved only by charge conjugation, and the defect $\mathcal{K}_{\omega}$ defect provided that $g=1$. 

While the exotic theory enjoys $SL(2,\mathbb{R})$ rotation of the gauge fields, a direct construction of the foliated defects has proven to be technically complicated, for a simple rotation of the gauge fields will break the foliation. We employ the exotic/foliated duality for expecting that the same symmetry and corresponding defects exist even in the foliated side. In the following, we construct the above defects from a path integral perspective. They are defined on a codim-1 surface $\Sigma$, which in this case allows for a discrete torsion term as in \cite{Paznokas:2025auw}. All gauge fields living on the defects take value in $\mathbb{R}$, as well as their lagrange multiplier.

\subsubsection*{$\mathcal{T}$ defect}

Let us support the defects on a closed surface $\Sigma = T^3$ defined at fixed $r = r_*$, so that the defect is parallel to the topological and physical boundaries of the SymTFT. The action for the $\mathcal{T}_l$ defect reads
\begin{equation}\label{eq:DefectT}
    S_l^{\mathcal{T}} [\Sigma] = \frac{i}{2\pi} \int_{\Sigma} v_t A_{xy} + v_{xy} A_t + v_t \partial_x \partial_y \varphi + v_{xy} \partial_t \varphi + \frac{1}{l^{xy}} v_t v_{xy} \; ,
\end{equation}
where the measure $dxdydt$ is understood, $v_t$ and $v_{xy}$ are gauge fields living on the defect, and $\varphi$ is a lagrange multiplier that ensures the flatness of the gauge fields $v$. The last term is a discrete torsion term, needed for realizing the action of $\mathcal{T}_l$ on the bulk gauge fields. Gauge-invariance is ensured on the e.o.m. and with $\delta \varphi = - \lambda$, $\delta v_t = \partial_t \lambda_v$ and $\delta v_{xy} = \partial_x \partial_y \lambda_v$.

The equation of motions of the gauge fields in presence of the defect read
\begin{align}\label{eq:eomGaugeDefect}
    &\partial_t \tilde{A}_r - \partial_r \tilde{A}_t = - v_{t} \delta(r - r_*) \; , \nonumber \\
    &\partial_x \partial_y \tilde{A}_r - \partial_r \tilde{A}_{xy} = - v_{xy} \delta(r - r_*) \; , \nonumber \\
    & v_t = - l^{xy} \left( A^{\Sigma}_t + \partial_t \varphi \right) \; , \nonumber \\
    & v_{xy} = - l^{xy} \left( A^{\Sigma}_{xy} + \partial_x \partial_y \varphi \right) \; ,
\end{align}

where $A^{\Sigma}$ is the value of on the defect. The bulk gauge fields are discontinuous from left to right of the defect and we take $A^{\Sigma} = \frac{1}{2} \left( A_L + A_R \right)$, similarly for the $\tilde{A}$ fields.

In order to see the action on the bulk gauge fields, we consider a small tube transverse to $\Sigma$ and extending from as $r_1 \leq r \leq r_2$, with $r_* \in [ r_1, r_2 ]$. By Stokes theorem, the integral in the tube of the e.o.m. yields 
\begin{align}\label{eq:ActionDefectT}    
& \int_{r_1}^{r_2} dr \oint_{\gamma(t)} dt \,  \left( \partial_r \tilde{A}_t - \partial_t \tilde{A}_r \right) = \oint_{\gamma(t),r_2} dt \tilde{A}_t - \oint_{\gamma(t),r_1} dt \tilde{A}_t \; , \nonumber \\[4pt]
& \int_{r_1}^{r_2} dr \oint_{\gamma(t)} dt \,  \left( - \partial_r A_t + \partial_t A_r \right) = - \oint_{\gamma(t),r_2} dt A_t + \oint_{\gamma(t),r_1} dt A_t \; ,
\end{align}
where $\gamma(t)$ is a closed line on the $t$ direction. By inserting \eqref{eq:eomGaugeDefect} and sending $r_1 \to r_*$ and $r_2 \to r_*$ from the left and from the right respectively, we get the expected action 
\begin{align}
    &\tilde{A}^R_t = \tilde{A}^L_t + l^{xy} \, A^{\Sigma}_t \; , \nonumber \\
    &A^R_t = A^L_t  \; , 
\end{align}
with $A^{R,L}$ the right and left value of the gauge field. This reproduces the action of the $\mathcal{T}_l$ defect in \eqref{eq:Iwasawa}. The argument for the remaining bulk gauge fields follows the same lines. 

The defect can be opened on the time direction, and the complete action with $\partial \Sigma \neq 0$ reads
\begin{align}
    S_l^{\mathcal{T}} [\Sigma] + \frac{i}{2 \pi} \int_{\partial \Sigma} \sigma A_{xy} + \sigma \partial_x \partial_y \varphi + \frac{1}{2l^{xy}} \left( \sigma \partial_x \partial_y \sigma + 2 \sigma v_{xy} \right) \; ,
\end{align}
where in the second term the measure $dxdy$ is understood, and with $\delta \sigma = - \lambda_v$.

\subsubsection*{$\mathcal{A}$ defect}

The action of $\mathcal{A}_s$ defect has the form
\begin{align}\label{eq:DefectA}
    S_{s}^{\mathcal{A}} [\Sigma] = & \frac{i}{2\pi} \int_{\Sigma} v_t A_{xy} + v_{xy} A_t + v_t \partial_x \partial_y \varphi + v_{xy} \partial_t \varphi + \tilde{v}_t \Tilde{A}_{xy} + \tilde{v}_{xy} \Tilde{A}_t + \tilde{v}_t \partial_x \partial_y \tilde{\varphi} + \tilde{v}_{xy} \partial_t \tilde{\varphi} \nonumber \\[4pt]
& + \frac{s + 1}{2(s-1)} \left( v_{t} \tilde{v}_{xy} + v_{xy} \tilde{v}_t \right) \nonumber \\[4pt]
& + \frac{i}{2\pi} \int_{\partial \Sigma} \sigma A_{xy} + \sigma \partial_ x \partial_y \varphi + \tilde{\sigma} \tilde{A}_{xy} + \tilde{\sigma} \partial_ x \partial_y \tilde{\varphi} + \frac{s + 1}{2(s-1)} \left( \sigma \partial_x \partial_y \tilde{\sigma} + \sigma \tilde{v}_{xy} + \tilde{\sigma} v_{xy} \right)  \; ,
\end{align}
where the gauge fields on the defect transform as $\delta \tilde{\varphi} = - \tilde{\lambda}$, $\delta \tilde{v}_t = \partial_t \tilde{\lambda}_v$ and $\delta \tilde{v}_{xy} = \partial_x \partial_y \tilde{\lambda}_v$, and $\delta \tilde{\sigma} = - \tilde{\lambda}_v$. This lagrangian reproduces the action on the bulk gauge fields of the $\mathcal{A}_s$ defect in \eqref{eq:Iwasawa}.

\subsubsection*{$\mathcal{K}$ defect}

The lagrangian form that reproduces the action of the $\mathcal{K}_{\omega}$ defect is
\begin{align}\label{eq:DefectK}
    S_{\omega}^{\mathcal{K}}[\Sigma] =& \frac{i}{2\pi} \int_{\Sigma} v_t A_{xy} + v_{xy} A_t + v_t \partial_x \partial_y \varphi + v_{xy} \partial_t \varphi + \tilde{v}_t \Tilde{A}_{xy} + \tilde{v}_{xy} \Tilde{A}_t + \tilde{v}_t \partial_x \partial_y \tilde{\varphi} + \tilde{v}_{xy} \partial_t \tilde{\varphi} \nonumber \\[4pt]
    &+ \frac{\mathbb{1}^{xy} \sin \omega }{2(\cos \omega - 1)} \left (v_{xy} v_t - \tilde{v}_{xy} \tilde{v}_t \right) \nonumber \\[4pt]
    &+ \frac{i}{2 \pi} \int_{\partial \Sigma} \sigma A_{xy} + \sigma \partial_x \partial_y \varphi + \tilde{\sigma} \tilde{A}_{xy} + \tilde{\sigma} \partial_x \partial_y \tilde{\varphi} \nonumber \\[4pt] 
    &+ \frac{\mathbb{1}^{xy} \sin \omega}{4(\cos \omega - 1)} \left( \sigma \partial_x \partial_y \sigma + \tilde{\sigma} \partial_x \partial_y \tilde{\sigma} + 2 \sigma v_{xy} + 2 \tilde{\sigma} \tilde{v}_{xy}  \right)\; .
\end{align}

Fixing the value of $g=1$, the $\mathcal{K}_{\omega}$ defect preserves the conformal boundary conditions. Thus, when $\partial \Sigma \neq 0$ the defect can end on the boundary of the SymTFT and it generates an $SO(2) \subset SL(2,\mathbb{R})$ which survives on the physical boundary. In order to characterize the boundary symmetry, we analyse the fusion of two such open defects $\mathcal{K}_{\omega}$, $\mathcal{K}_{-\omega}$, see App. \ref{appx:Fusion} for the detailed computations. We find that such a fusion yields
\begin{equation}
    S^{\mathcal{K}\mathcal{K}^{-1}}=\frac{i}{2\pi}\int_{\partial\Sigma}\eta A_{xy}+\tilde{\eta}\tilde{A}_{xy}+\eta\partial_x\partial_y\varphi^{+}+\tilde{\eta}\partial_x\partial_y\tilde{\varphi}^{+} \; ,
\end{equation}
where to compute the fusion is computed in the basis $v^{\pm} = v \pm w$, $\varphi^{\pm} = \varphi \pm \psi$, with $v$ and $w$ the gauge fields on the two defects, $\varphi$ and $\psi$ their lagrange multiplier, and $\eta = \partial_t v^+$. The result of this fusion is not trivial, and it implements the higher gauging of  $\mathbb{R} \times \mathbb{R}$ symmetry on the boundary $\partial \Sigma$. Upon slab compactification, this generates a continuous non-invertible symmetry $SO(2)$ for the XY-plaquette model with no conditions on the coupling. In case the $\theta$-term discussed in sec. \ref{sec:exotictheta} is present, the bulk theory and the condensation defects are not changed, hence the continuous non-invertible symmetry hols at any $R$ and $\theta$. Continuing the similarity with Maxwell, the same continuous symmetry arises with a similar mechanism in the case of Maxwell theory in 3+1 dimensions \cite{Paznokas:2025auw}.

\subsection{Condensation defects in 4+1d and XYZ cube}

The 0-form symmetries of the exotic theory in 4+1 dimensions
\begin{equation}
    \cL= \frac{i}{2\pi} \left[ \Tilde{A}_t \left( \partial_r A_{xyz} - \partial_x \partial_y \partial_z A_r \right) + \Tilde{A}_r \left( \partial_x \partial_y \partial_z A_t - \partial_t A_{xyz} \right) + \Tilde{A}_{xyz} \left( \partial_t A_r - \partial_r A_t \right) \right] \; ,
\end{equation}
are represented by a rescaling of the gauge field, given by the matrix
\begin{equation}
    \mathcal{A}_s = 
    \begin{pmatrix}
    s & 0 \\
    0 & \frac{1}{s}
    \end{pmatrix} \; ,
\end{equation}
and the transformation that exchanges the gauge fields as $A \to \tilde{A}$ and $\tilde{A} \to A$, given by the matrix
\begin{equation}
    \mathcal{K} = 
    \begin{pmatrix}
    0 & 1 \\
    1 & 0
    \end{pmatrix} \; ,
\end{equation}
and it leaves the conformal boundary conditions in eq. \eqref{eq:XYZconformalBC} invariant if $g=1$. At such value, the $\mathcal{K}$ defect can end on the boundary. It is interesting to notice that the symmetries of the exotic theory in 4+1d are smaller than the previous model in 3+1. Consequently, the XYZ cube obtained after slab compactification does not possess the $SO(2)$ that survive for the XY plaquette, but rather a discrete symmetry that exchanges the two fields $A$ and $\tilde{A}$. 

Supported on a codim-1 surface $\Sigma$ defined at $r=r_*$, the action of the defect $\mathcal{K}$ has the form
\begin{align}\label{eq:XYZDefectK}
    S^{\mathcal{K}}[\Sigma] &= \frac{i}{2 \pi} \int_{\Sigma} v_t \left( A_{xyz} - \mathbb{1}^{xyz}\tilde{A}_{xyz} \right) - v_{xyz} \left( A_t - \mathbb{1}^{xyz}\tilde{A}_t \right) - v_t \partial_x \partial_y \partial_z \varphi + v_{xyz} \partial_t \varphi \nonumber \\[4pt]
    &+ \frac{i}{2 \pi} \int_{\partial \Sigma} \sigma \left( A_{xyz} - \mathbb{1}^{xyz}\tilde{A}_{xyz} \right) + \varphi \partial_x \partial_y \partial_z \sigma  \; ,
\end{align} 
where $v_t$ and $v_{xyz}$ are gauge field living on the defect, whose flatness is ensured by $\varphi$ and $\chi$, where $\sigma$ is defined on the boundary $\partial \Sigma$ when the defect is open. Gauge invariance is ensured with $\delta v_t = \partial_t \lambda_v$, $\delta v_{xyz} = \partial_x \partial_y \partial_z \lambda_v$, $\delta \varphi = \lambda -\tilde{\lambda}$, and $\delta \sigma = - \lambda_v$. 

The fusion of two such defects is non-trivial, see app. \ref{appx:Fusion} for the explicit computation. In facts, when the boundary $\partial \Sigma \neq 0$ we get
\begin{equation}
    S^{\mathcal{K}\mathcal{K}}=\frac{i}{2 \pi} \int_{\partial \Sigma} \sigma'' \left( A_{xyz} - \mathbb{1}^{xyz}\tilde{A}_{xyz} \right) + \varphi \partial_x \partial_y \partial_z \sigma'' + \varphi' \partial_x \partial_y \partial_z \sigma'  \; .
\end{equation}
By compactification, this generate a discrete non-invertible symmetry for the XYZ-cube, in agreement with the interface discussed in \cite{Spieler:2024fby}.

\section{Conclusions and outlook}\label{sec:conclusion}

In this work, we study the symmetries and dualities of the XY plaquette and XYZ cube models, using the framwork of \textit{Mille-feuille} SymTFT developed in \cite{Apruzzi:2025mdl}, and constructing codim-1 condensation defects in the bulk. For the XYZ cube model, we find the interfaces of \cite{Spieler:2024fby, Gorantla:2020xap}. On the other hand, the XY plaquette model \cite{Seiberg:2020bhn} is more interesting. The bulk of the SymTFT is similar to the SymTFT bulk for Maxwell theory, since it has an $SL(2,\mathbb{R})$ symmetry. This goes beyond a mere resemblance, as we show that the XY plaquette admits an exotic $\theta$-term, and it enjoys a continuous non-invertible duality symmetry $SO(2)$ at any value of the couplings $R$ and $\theta$. 

These results suggest some directions for further research. The XY-plaquette is analysed in \cite{Seiberg:2020bhn} starting from a lattice description and studying the effective low energy theory. In our work, the $\theta$-term is added to the model in continuum limit. A lattice derivation along the lines of \cite{Seiberg:2020bhn} would clarify its UV origin. Regarding foliated descriptions, describing the condensation defects by summing over all possible insertion of foliated defects requires a consistent measure to be defined, as highlighted in Section \ref{sec:CondDefects}. Furthermore, the duality symmetries described mix the discrete rotational symmetry and charge conjugation. It would be interesting to describe this point by employing a SymTFT that includes spacetime symmetries \cite{Apruzzi:2025hvs, Pace:2025hpb}.

\section*{Acknowledgement}

The authors want to thank Fabio Apruzzi for discussion and a careful reading of the manuscript; and Riccardo Argurio, Jeremias Aguilera Damia, Giovanni Galati and Stathis Vitouladitis for discussion. The work of FB is supported in part by the Italian MUR Departments of Excellence grant 2023-2027 "Quantum Frontiers”. The work of SM is supported by the University of Padua under the 2023 STARS Grants@Unipd programme (GENSYMSTR – Generalized Symmetries from Strings and Branes) and in part by the Italian MUR Departments of Excellence grant 2023-2027 "Quantum Frontiers”.

\appendix

\section{Fusion of condensation defects}\label{appx:Fusion}

\subsection*{Defects in 3+1 SymTFT} 

The defect $\mathcal{K}_{\omega}$ in eq. \eqref{eq:DefectK} can end on the boundary of the SymTFT. The fusion of two such open defects $\mathcal{K}_{\omega}$, $\mathcal{K}_{-\omega}$ reads
\begin{equation}
\begin{split}
    S^{\mathcal{K}\mathcal{K}^{-1}} = & \frac{i}{2\pi}\int_\Sigma(v_t+w_t)A_{xy}+(v_{xy}+w_{xy})A_t+(\tilde{v}_t+\Tilde{w}_t)\Tilde{A}_{xy}+(\tilde{v}_{xy}+\Tilde{w}_{xy})\Tilde{A}_t\\[4pt]
    &+v_t\partial_x\partial_y\varphi+v_{xy}\partial_t\varphi+\tilde{v}_t\partial_x\partial_y\Tilde{\varphi}+\tilde{v}_{xy}\partial_t\Tilde{\varphi}+w_{t}\partial_x\partial_y\psi+w_{xy}\partial_t\psi\\[4pt]
    &+\Tilde{w}_t\partial_x\partial_y\Tilde{\psi}+\Tilde{w}_{xy}\partial_t\Tilde{\psi}+\frac{1}{2}(v_{xy}\Tilde{w}_t+v_{t}\Tilde{w}_{xy}-w_{xy}\tilde{v}_t-w_{t}\tilde{v}_{xy})\\[4pt]
    &\frac{ \mathbb{1}^{xy} \sin \omega}{2(\cos \omega-1)}(v_{xy}v_t - \tilde{v}_{xy}\tilde{v}_t - w_{xy}w_t + \Tilde{w}_{xy}\Tilde{w}_{t})\\[4pt]
    &+\frac{i}{2\pi}\int_{\partial\Sigma}(\sigma+\rho) A_{xy}+(\Tilde{\sigma}+\Tilde{\rho})\Tilde{A}_{xy}+\sigma\partial_x\partial_y\varphi+\Tilde{\sigma}\partial_x\partial_y\Tilde{\varphi}+\rho\partial_x\partial_y\psi+\Tilde{\rho}\partial_x\partial_y\Tilde{\psi}\\[4pt]
    &+\frac{\mathbb{1}^{xy} \sin \omega}{4(\cos \omega-1)}\left(\sigma\partial_x\partial_y\sigma+\Tilde{\sigma}\partial_x\partial_y\Tilde{\sigma}+2\sigma v_{xy}+2\Tilde{\sigma}\tilde{v}_{xy}\right.\\[4pt] &\left. -\rho\partial_x\partial_y\rho - \Tilde{\rho}\partial_x\partial_y\Tilde{\rho} - 2\rho w_{xy} - 2\Tilde{\rho}\Tilde{w}_{xy}\right) \; ,
\end{split}
\end{equation}
where we are also including the contribution given by the fields between the defects \cite{Antinucci:2022vyk, Argurio:2024ewp, Paznokas:2025auw}. By the following field redefinitions
\begin{equation}
    \begin{split}
        &v_t^\pm=v_t\pm w_t \; , \qquad v_{xy}^\pm=v_{xy}\pm w_{xy} \; , \\[4pt]
        &\tilde{v}_t^\pm=\tilde{v}_t\pm\Tilde{w}_{t} \; , \qquad \tilde{v}_{xy}^\pm=\tilde{v}_{xy}\pm\Tilde{w}_{xy} \; , \\[4pt]
        &\varphi^{\pm}=\varphi\pm\psi \; , \qquad \sigma^{\pm}=\sigma\pm\rho \; , \\[4pt]
        &\tilde{\varphi}^{\pm}=\tilde{\varphi}\pm\tilde{\psi} \; , \qquad \tilde{\sigma}^{\pm}=\tilde{\sigma}\pm\tilde{\rho} \; ,
    \end{split}
\end{equation}
we can write 
\begin{equation}
\begin{split}
    S^{\mathcal{K}\mathcal{K}^{-1}}& = \frac{i}{2\pi}\int_\Sigma v_t^+A_{xy}+v_{xy}^+A_t+\tilde{v}_t^+\Tilde{A}_{xy}+\tilde{v}_{xy}^+\Tilde{A}_t\\[4pt]
    &+\frac{1}{2}(v_t^+\partial_x\partial_y\varphi^++v_{xy}^+\partial_t\varphi^++v_t^-\partial_x\partial_y\varphi^-+v_{xy}^-\partial_t\varphi^-+\tilde{v}_t^+\partial_x\partial_y\Tilde{\varphi}^++\tilde{v}_{xy}^+\partial_t\Tilde{\varphi}^+ \\[4pt]
    &+ \tilde{v}_t^-\partial_x\partial_y\Tilde{\varphi}^-+\tilde{v}_{xy}^-\partial_t\Tilde{\varphi}^-)
    +\frac{1}{4}(v_{xy}^+\Tilde{v}^-_t+v_{t}^+\Tilde{v}^-_{xy}-v^-_{xy}\Tilde{v}^+_t-v^-_{t}\Tilde{v}^+_{xy})\\[4pt]
    &\frac{\mathbb{1}^{xy}\sin \omega}{4(\cos \omega -1)}( v_{xy}^+v_t^-+v_t^+v_{xy}^- - \tilde{v}_{xy}^+\tilde{v}_t^- - \tilde{v}_t^+\tilde{v}_{xy}^-)\\[4pt]
    &+\frac{i}{2\pi}\int_{\partial\Sigma}\sigma^+ A_{xy}+\Tilde{\sigma}^+\Tilde{A}_{xy}+\frac{1}{2}(\sigma^+\partial_x\partial_y\varphi^++\sigma^-\partial_x\partial_y\varphi^-+\Tilde{\sigma}^+\partial_x\partial_y\Tilde{\varphi}^++\Tilde{\sigma}^-\partial_x\partial_y\Tilde{\varphi}^-)\\[4pt]
    &\frac{\mathbb{1}^{xy}\sin \omega}{4(\cos \omega -1)}(\sigma^+\partial_x\partial_y\sigma^-+\Tilde{\sigma}^+\partial_x\partial_y\Tilde{\sigma}^-+2\sigma^+ v_{xy}^-+2\sigma^- v_{xy}^++2\Tilde{\sigma}^+\tilde{v}_{xy}^-+2\Tilde{\sigma}^-\tilde{v}_{xy}^+) \; .
\end{split}
\end{equation}
Integrating out the gauge fields $v^-$ and $\Tilde{v}^-$ we get
\begin{equation}
    \begin{split}
        &\partial_t\varphi^-=-\frac{\mathbb{1}^{xy}\sin{\omega}}{2(\cos \omega -1)}v_t^++\frac{1}{2}\Tilde{v_{t}}^+ \; , \\[4pt]
        &\partial_t\Tilde{\varphi}^-=+\frac{\mathbb{1}^{xy}\sin{\omega}}{2(\cos \omega-1)}\tilde{v}_t^+-\frac{1}{2}v_{t}^+ \; , \\[4pt]
        &\partial_x\partial_y\varphi^-=-\frac{\mathbb{1}^{xy}\sin{\omega}}{2(\cos \omega-1)}v_{xy}^{+}+\frac{1}{2}\tilde{v}_{xy}^{+} \; , \\[4pt]
        &\partial_x\partial_y\tilde{\varphi}^-=+\frac{\mathbb{1}^{xy}\sin{\omega}}{2(\cos \omega -1)}\tilde{v}_{xy}^{+}-\frac{1}{2}v_{xy}^{+} \; ,\\[4pt]
    \end{split}
\end{equation}
furthermore, the eoms of $v^-$ and $\Tilde{v}^-$ on the boundary set $\sigma^+$ and $\tilde{\sigma}^{+}$ to zero, thus trivializing the edge mode. We can now define:
\begin{equation}\label{eq:etadef}
    \begin{split}
        &v_t^+=\partial_t\eta = 2 \partial_t \left( - \frac{\mathbb{1}^{xy}}{\nu - \nu^{-1}}\varphi^- + \frac{1}{\nu^2-1}\tilde{\varphi}^- \right)\\[4pt]
        &v_{xy}^+=\partial_x\partial_y\eta= 2\partial_x\partial_y \left(-\frac{\mathbb{1}^{xy}}{\nu-\nu^{-1}}\varphi^- + \frac{1}{\nu^2-1}\tilde{\varphi}^- \right)\\[4pt]
        &\tilde{v}_t^+=\partial_t\tilde{\eta}= 2\partial_t \left(-\frac{1}{\nu^2-1}\varphi^- +\frac{\mathbb{1}^{xy}}{\nu-\nu^{-1}}\tilde{\varphi}^- \right)\\[4pt]
        &\tilde{v}_{xy}^+=\partial_x\partial_y\tilde{\eta}=2\partial_x\partial_y \left(-\frac{1}{\nu^2-1}\varphi^- +\frac{\mathbb{1}^{xy}}{\nu-\nu^{-1}}\tilde{\varphi}^-\right)\\[4pt]
    \end{split}
\end{equation}
with $ \nu=\frac{\sin \omega}{\cos \omega -1}$. Using eq. \eqref{eq:etadef} and the bulk eoms, the action on $\Sigma$ trivializes and we are left only with the boundary $\partial \Sigma$ action. Integrating out the auxiliary fields $\varphi^-$ and $\tilde{\varphi}^-$, we get
\begin{equation}
    S^{KK^{-1}}=\frac{i}{2\pi}\int_{\partial\Sigma}\eta A_{xy}+\tilde{\eta}\tilde{A}_{xy}+\eta\partial_x\partial_y\varphi^{+}+\tilde{\eta}\partial_x\partial_y\tilde{\varphi}^{+} \; ,
\end{equation}
which implements the same higher gauging of the $\mathbb{R} \times \mathbb{R}$ on the boundary $\partial \Sigma$. 

\subsection*{Defects in 4+1 SymTFT}

The defect $\mathcal{K}$ in eq. \eqref{eq:XYZDefectK} can end on the boundary of the SymTFT. We compute the fusion of two such defects as
\begin{align}
    S^{\mathcal{K}\mathcal{K}}&= \frac{i}{2 \pi} \int_{\Sigma} (v_t+v'_t) \left( A_{xyz} - \mathbb{1}^{xyz}\tilde{A}_{xyz} \right) - (v_{xyz}+v'_{xyz}) \left( A_t - \mathbb{1}^{xyz}\tilde{A}_t \right)-v_{xyz}v'_{t}+v_{t}v'_{xyz}\nonumber\\[4pt]
    &+ \frac{i}{2 \pi} \int_{\Sigma}  + v_{xyz} \partial_t \varphi - v_t \partial_x \partial_y \partial_z \varphi + v'_{xyz} \partial_t \varphi' - v'_t \partial_x \partial_y \partial_z \varphi' \nonumber \\[4pt]
    &+ \frac{i}{2 \pi} \int_{\partial \Sigma} (\sigma+\sigma') \left( A_{xyz} - \tilde{A}_{xyz} \right) + \varphi \partial_x \partial_y \partial_z \sigma + \varphi' \partial_x \partial_y \partial_z \sigma'  \; ,
\end{align}
where we included the contribution given by the fields between the defects. With the appropriate field redefinitions
\begin{align}
    \varphi&=\varphi'-\varphi \; , \nonumber\\[4pt]
    \sigma&=\sigma'+\sigma \; , \nonumber\\[4pt]
    v_t&=v_t+v'_t\;,\;\;v_{xyz}^+=v_{xyz}+v'_{xyz} \; ,
\end{align}
we get
\begin{align}
    S^{\mathcal{K}\mathcal{K}}&= \frac{i}{2 \pi} \int_{\Sigma} v_t \left( A_{xyz} - \tilde{A}_{xyz} \right) - v_{xyz} \left( A_t - \tilde{A}_t \right)-v_{xyz}v'_{t}+v_{t}v'_{xyz}\nonumber\\[4pt]
    &+ \frac{i}{2 \pi} \int_{\Sigma}  + v_{xyz} \partial_t \varphi - v_t \partial_x \partial_y \partial_z \varphi + v'_{xyz} \partial_t \varphi' - v'_t \partial_x \partial_y \partial_z \varphi' \nonumber \\[4pt]
    &+ \frac{i}{2 \pi} \int_{\partial \Sigma} \sigma \left( A_{xyz} - \mathbb{1}^{xyz}\tilde{A}_{xyz} \right) + \varphi \partial_x \partial_y \partial_z \sigma + \varphi' \partial_x \partial_y \partial_z \sigma'  \; .
\end{align}
If we then substitute the eoms of the field $v_t'$ and $v'_{xyz}$
\begin{align}    
     &v_{xyz}=\partial_x\partial_y\partial_z\varphi'\; , \nonumber \\[4pt] 
     &v_t=\partial_t\varphi' \; ,
\end{align}
the bulk action trivializes and we find
\begin{equation}
    S^{\mathcal{K}\mathcal{K}}=\frac{i}{2 \pi} \int_{\partial \Sigma} \sigma'' \left( A_{xyz} - \mathbb{1}^{xyz}\tilde{A}_{xyz} \right) + \varphi \partial_x \partial_y \partial_z \sigma'' + \varphi' \partial_x \partial_y \partial_z \sigma'  \; ,
\end{equation}
where we redefined $\sigma''=\sigma+\varphi'$. 

\bibliography{Symbib}
\bibliographystyle{ytphys}
\end{document}